\documentclass[english,amsmath,amssymb,superscriptaddress,nofootinbib,twocolumn]{revtex4-1}
\usepackage[T1]{fontenc}
\usepackage[latin9]{inputenc}
\usepackage{amsmath}
\usepackage{amssymb}
\usepackage{graphicx}
\usepackage{babel}
\usepackage{color}
\usepackage[normalem]{ulem}

\begin{document}

\title{ Magnetic Field-controlled Transmission and Ultra-narrow Superradiant Lasing by Dark Atom-light Dressed States}

\author{Yuan Zhang}
\email{yzhuaudipc@zzu.edu.cn}

\address{School of Physics and Microelectronics, Zhengzhou University, Daxue Road 75,
 Zhengzhou 450052 China}

\address{Donostia International Physics Center, Paseo Manuel de Lardizabal
4, 20018 Donostia-San Sebastian (Gipuzkoa), Spain}

\author{Chong Xin Shan}
\address{School of Physics and Microelectronics, Zhengzhou University, Daxue Road 75,
 Zhengzhou 450052 China}

\author{Klaus M{\o}lmer}
\email{moelmer@phys.au.dk}

\address{Department of Physics and Astronomy, Aarhus University, Ny Munkegade
120, DK-8000 Aarhus C, Denmark}
\begin{abstract}
Optical lattice clock systems with ultra-cold strontium-88 atoms have been used to demonstrate superradiant lasing and magnetic field-controlled optical transmission. We explain these phenomena theoretically with a rigorous model for three-level atoms coupled to a single cavity mode. We identify a class of dark atom-light dressed states which become accessible due to mixing with bright dressed states in the presence of a magnetic field. We predict that these states, under moderate incoherent pumping, lead to lasing with a linewidth of only tens of Hz,  orders of magnitude smaller than the cavity linewidth and the atomic incoherent decay and pumping rates.
\end{abstract}
\maketitle

\paragraph{Introduction}

\begin{figure}[!ht]
\begin{centering}
\includegraphics[scale=0.55]{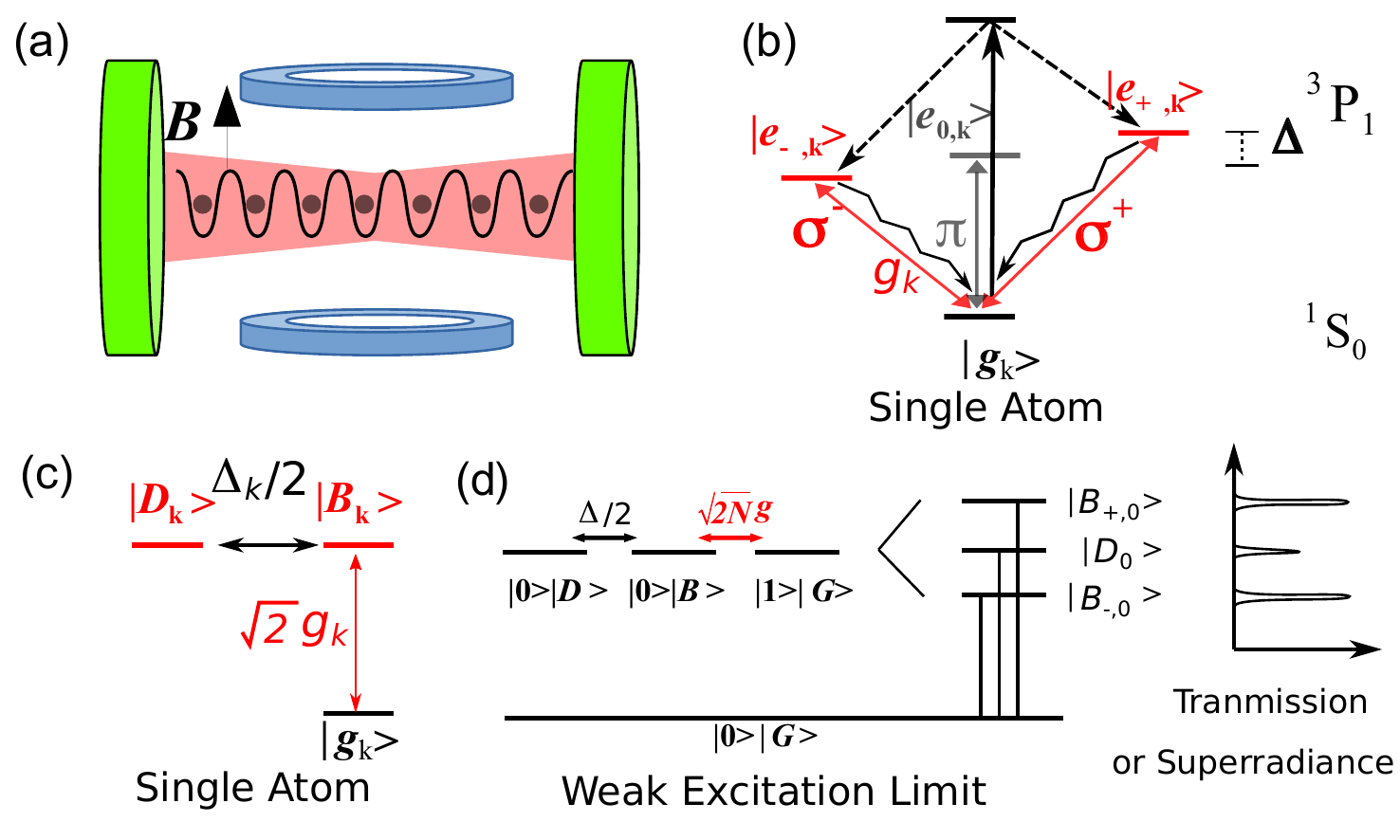}
\par\end{centering}
\caption{\label{fig:system} Panel (a) illustrates thousands of ${}^{88} {\rm Sr}$ atoms trapped in a one-dimensional optical lattice inside
an optical cavity. Panel (b) shows the ground state ($^{1}S_{0}$) and the three excited states ($^{3}P_{1}$), which are split due to the magnetic field. The atomic excitation is pumped coherently to a higher excited state (black solid arrow) and decays rapidly to the excited states of interest (dashed arrows) and returns to the ground state by spontaneous emission (wavy arrows) and coherent coupling to the cavity mode (red thin lines) \citep{MANorcia}. Panel (c) shows the equivalent scheme with one bright and one dark excited state in a single atom. Panel (d) shows that three dressed atom-light states are formed due to the Zeeman splitting and the collective atom-light interaction, leading to three peaks in the transmission spectrum and steady-state superradiance.}
\end{figure}

Conventional lasers rely on optical coherence established by
stimulated emission from a population-inverted medium
 and have a spectral linewidth set by the Schawlow-Townes limit
\cite{ALSchawlow}. In contrast, superradiant lasers rely on
coherence in the medium, established by the collective atom-light interaction, and have a minimal linewidth given by the Purcell enhanced atomic decay rate $\Gamma_c$ \cite{DMeiser,DMeiser1,JGBohnet,MANorciaSciAdv}.
Recent experiments \citep{MANorcia} showed that the two lasing mechanisms may co-exist in a superradiant crossover regime with, e.g, ${}^{88} {\rm Sr}$ alkaline earth atoms trapped in a one-dimensional optical lattice inside an optical cavity, see Fig. \ref{fig:system}(a). Theoretical studies \cite{KDebnath,DATieri} revealed that lasing in this regime benefits from both the optical and atomic coherence and can achieve a linewidth even smaller than $\Gamma_c$.

The same system was employed recently \citep{MNWinchester} to demonstrate magnetic field-controlled transmission of light, which was explained with a model involving three coupled oscillators \citep{MNWinchester,ZXLiu}. In this Letter, we develop a more rigorous description in the spirit of the Jaynes-Cummings model for two-level systems \citep{Jaynes}, to explain this phenomenon with dressed atom-light states, see Fig.\ref{fig:system} (c) and (d). Our analysis further reveals that the interplay  of bright and dark dressed states is responsible for a new lasing mechanism with ultra-narrow linewidth.

\paragraph{Model} Fig. \ref{fig:system} (b) shows  the singlet ground state $^{1}S_{0}$, the triplet excited state $^{3}P_{1}$ and a further excited state used for incoherent excitation of the atomic system. The spin-forbidden $^{1}S_{0}-{}^{3}P_{1}$ transition becomes allowed in ${}^{88} {\rm Sr}$ atoms due to the spin-orbit interaction induced state-mixing \citep{MMBoyd}. In the presence of a static magnetic field, the triplet excited state of the $k^{th}$ atom is split due to the Zeeman effect into three states denoted as $\left|e_{\pm,k}\right\rangle,\left|e_{0,k}\right\rangle$ and the atom emits  $\sigma_{\pm},\pi$ polarized light at different frequencies. The transitions  $\sigma_{\pm}$ both couple to the single fundamental cavity mode with polarization in the  horizontal direction.

To analyze the scenario detailed above, the $N$ atoms are described by the Hamiltonian $H_{a}=\sum_{k=1}^{N}H_{k}$ with the single atom contribution $H_{k}= \sum_{s=e_{+},e_{-}}\hbar\omega_{s}\left|s_{k}\right\rangle \left\langle s_{k}\right|$, where $\omega_{e_{\pm}} = \omega_{a,k} \pm \Delta_k/2$. Here, the energy of the ground state is set as zero, $\omega_{a,k}=2\pi c/\lambda$ is the frequency of the $^{1}S_{0}-{}^{3}P_{1}$ transition with the wavelength $\lambda=689$ nm, the splitting $\Delta_k=2\pi\times2.1B$ MHz is determined by the static magnetic field $B$ in unit of Gauss \citep{MNWinchester}.  The cavity mode is described by the Hamiltonian $H_{c}=\hbar\omega_{c}a^{+}a$ with the frequency $\omega_{c}$, and creation and annihilation operators
$a^{+}$ and $a$. The interaction between the atoms and the cavity mode is described by the Hamiltonian $H_{a-c}= \sum_{k} H_{a-c,k}$ with $H_{a-c,k}= \hbar g_k \left(\left|e_{+,k}\right\rangle \left\langle g_{k}\right|+ \left|e_{-,k}\right\rangle \left\langle g_{k}\right|\right)a+\mathrm{h.c.}$ in the rotating wave approximation, where we assume the same coupling strength $g_{k}=2\pi\times7.5$ kHz for both $\sigma_{\pm}$ transitions. We ignore the excited state $\left|e_{0,k}\right\rangle$ since it does not couple with the cavity mode of interest.

To understand the interaction between a  single atom and the cavity mode,  we introduce two new excited states $\left|B_{k}\right\rangle =\left(\left|e_{+,k}\right\rangle +\left|e_{-,k}\right\rangle \right)/\sqrt{2}$, $\left|D_{k}\right\rangle =\left(\left|e_{+,k}\right\rangle -\left|e_{-,k}\right\rangle \right)/\sqrt{2}$. The single atom-cavity mode interaction Hamiltonian $H_{a-c,k}  =\hbar \sqrt{2}g_{k} \left|B_{k}\right\rangle \left\langle g_{k}\right| a+h.c.$ indicates that the bright (dark) states $\left|B_{k}\right\rangle$ ($\left|D_{k}\right\rangle$) do (do not) couple with the cavity mode. But they are coupled by the magnetic field-induced Zeeman-splitting via $H_{a,k}  = \hbar \omega_{a} \left(\left|B_{k}\right\rangle \left\langle B_{k}\right|+\left|D_{k}\right\rangle \left\langle D_{k}\right|\right) +\hbar(\Delta_k/2)\sum_{k}\left(\left|B_{k}\right\rangle \left\langle D_{k}\right|+\left|D_{k}\right\rangle \left\langle B_{k}\right|\right)$, see Fig. \ref{fig:system} (c).

To describe the interaction between the entire atomic ensemble and the cavity mode, we consider the weak excitation limit and introduce the ground state $\left|G\right\rangle =\prod_{k}\left|g_{k}\right\rangle $ and collective singly-excited bright $\left|B\right\rangle   = N^{-1/2} \sum_{k}\left|B_{k}\right\rangle \prod_{j\neq k}\left|g_{k}\right\rangle$ and dark state  $\left|D\right\rangle = N^{-1/2}\sum_{k}\left|D_{k}\right\rangle \prod_{j\neq k}\left|g_{k}\right\rangle$. This allows us to approximate the interaction and atomic Hamiltonian as $H_{a-c} \approx\hbar\sqrt{2N}g \left|B\right\rangle \left\langle G\right|a+h.c.$, $H_{a} \approx \hbar\omega_{a}\left(\left|B\right\rangle \left\langle B\right|+\left|D\right\rangle \left\langle D\right|\right) +\hbar(\Delta/2)\left(\left|B\right\rangle \left\langle D\right|+\left|D\right\rangle \left\langle B\right|\right)$. Here, we assume identical atoms, i.e. $\omega_{a}=\omega_{a,k},\Delta = \Delta_{a,k}, g=g_k$ for all $k$.

\begin{figure}
\begin{centering}
\includegraphics[scale=0.27]{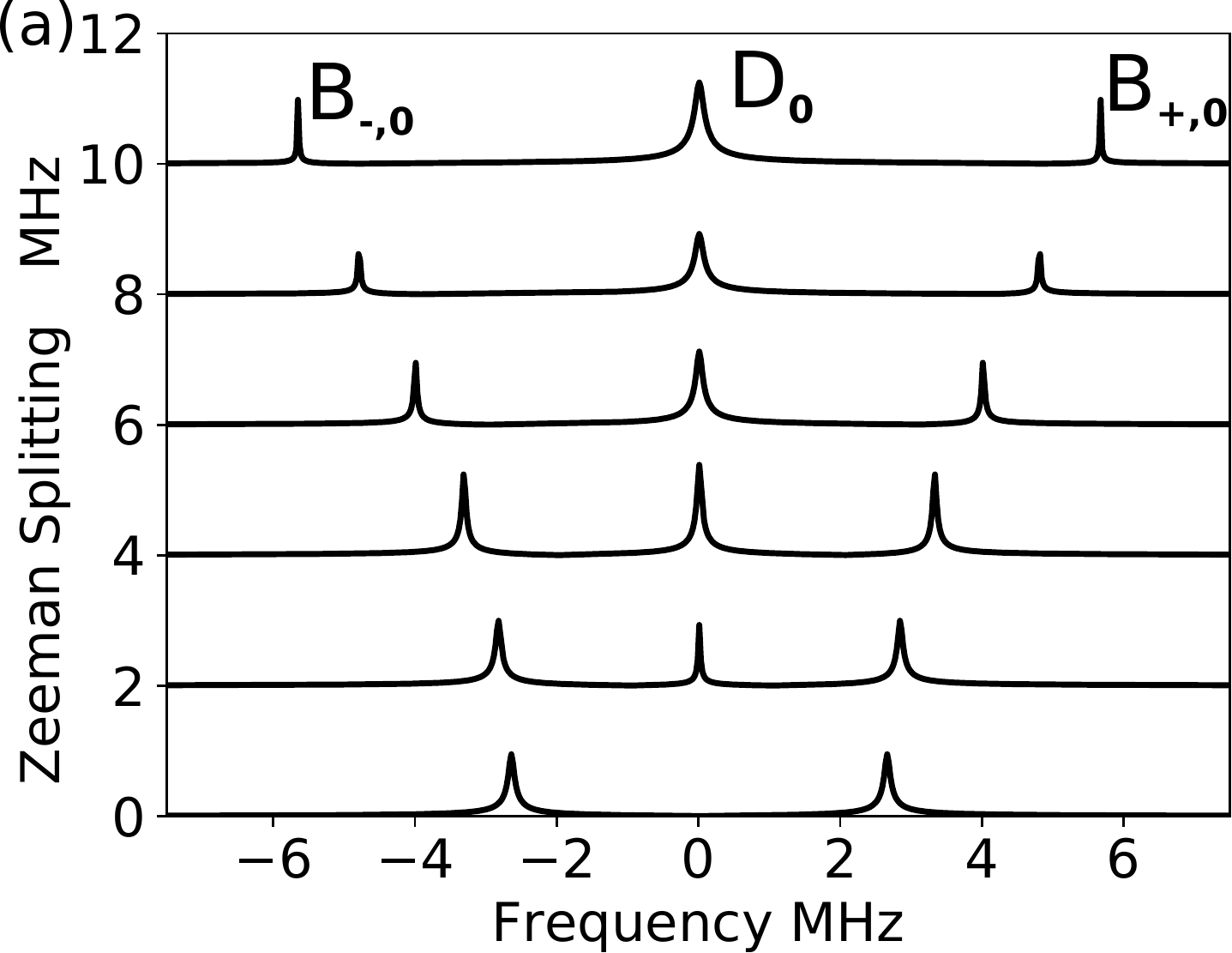}
\includegraphics[scale=0.26]{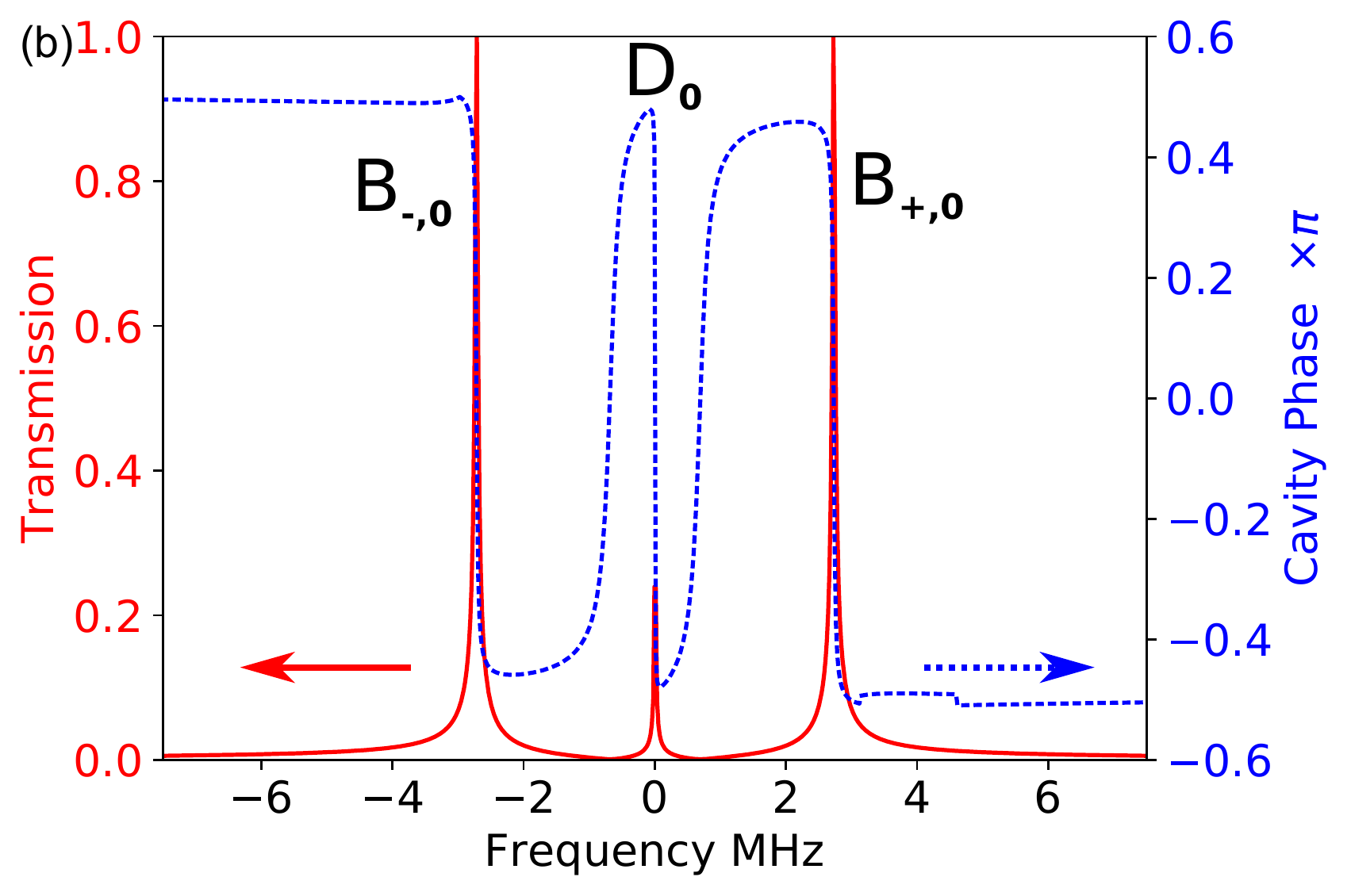}
\includegraphics[scale=0.27]{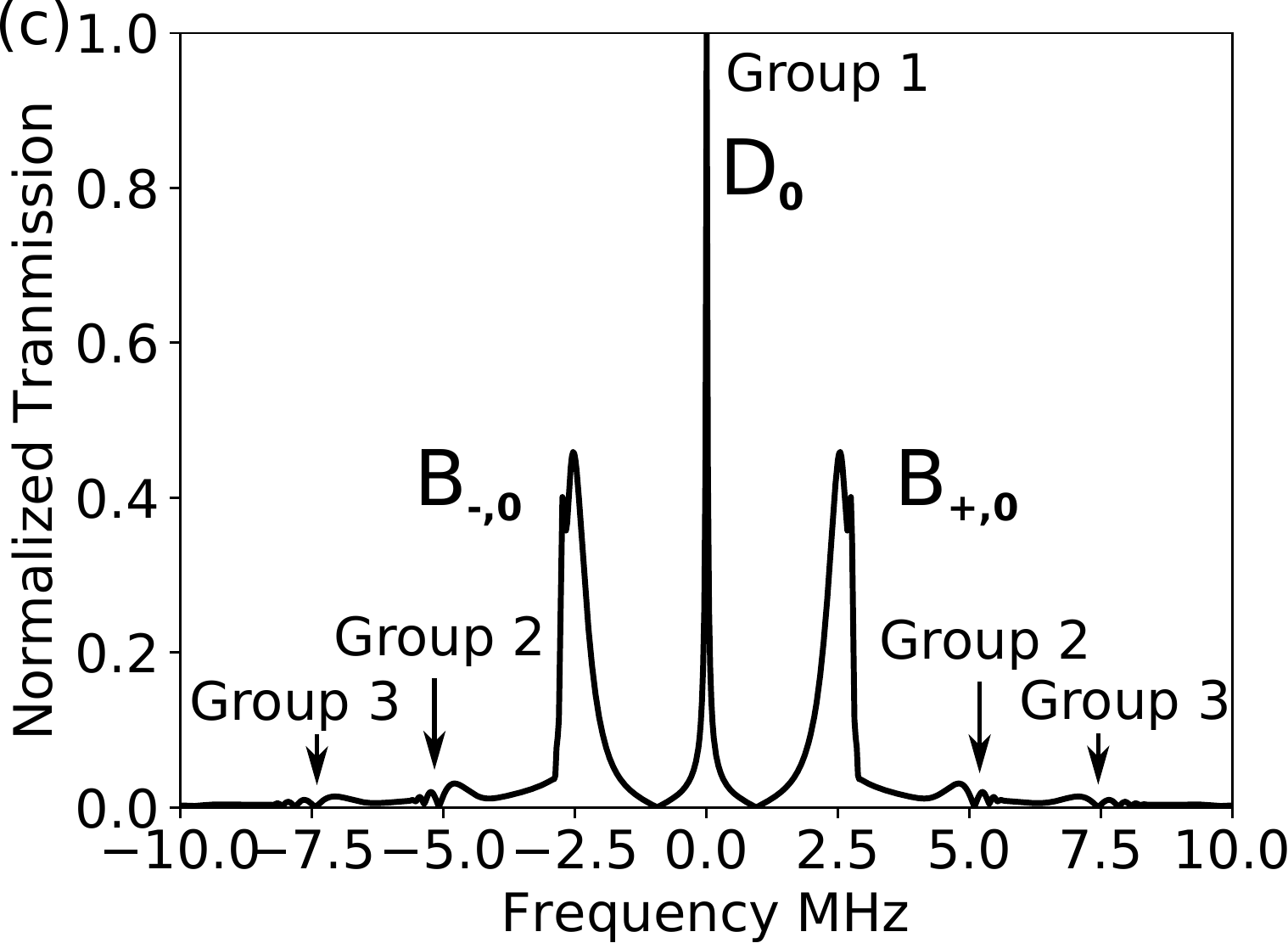}
\includegraphics[scale=0.27]{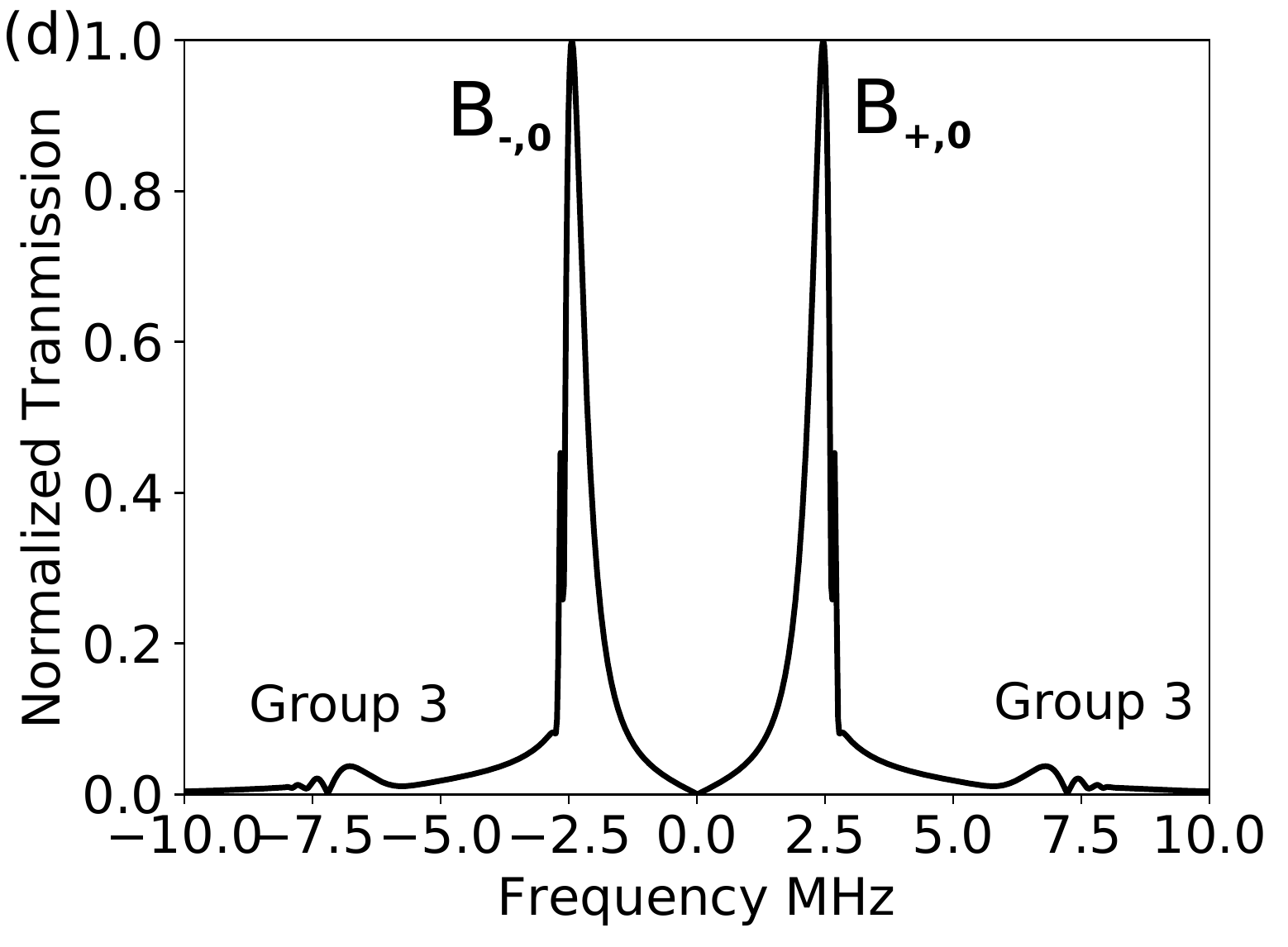}
\par\end{centering}
\caption{\label{fig:transmission} Magnetic field-controlled transmission spectrum (relative to $\omega_c$) of systems with $6.25\times10^{4}$ ${}^{88} {\rm Sr}$ atoms probed with a weak laser (a,b) and a strong laser (c,d). Panel (a) shows the transmission for different values of  Zeeman splitting $\Delta/2\pi$. Panel  (b) shows the transmission (solid black curve) and phase (dashed blue curve) for a given Zeeman splitting $\Delta/2\pi=2$ MHz. Panels  (c) and (d) show the transmission with and without Zeeman splitting $\Delta/2\pi=2$ MHz, respectively. The symbols $D_0,B_{\pm,0}$ and group 1,2,3 associate the peaks with the transitions between the dressed states as shown in Fig. \ref{fig:system}(d) and Fig. \ref{fig:dressed-states} in the Appendix \ref{sec:diag}. Other parameters are specified in the text. }
\end{figure}

In the spirit of the Jaynes-Cumming model we introduce the basis of product states $\left|n\right\rangle \left|D\right\rangle $, $\left|n\right\rangle \left|B\right\rangle $, $\left|n+1\right\rangle \left|G\right\rangle $ with the photon number states $\left|n\right\rangle $ ($n=0,1,...$) and the atom-ensemble states, and decompose
$H_{s} = H_c + H_a + H_{a-c}\approx\oplus_{n}H^{\left(n\right)}$ in this basis with
\begin{equation}
H^{\left(n\right)}=\hbar\left[\begin{array}{ccc}
\omega_{a} - \omega_c & \Delta/2 & 0\\
\Delta/2 & \omega_{a} - \omega_c & g_{n}\\
0 & g_{n} & 0
\end{array}\right]+\hbar\left(n+1\right)\omega_{c}.\label{eq:decomposedHamiltonian}
\end{equation}
Here, $g_{n}=\sqrt{n+1}\sqrt{2N}g$ is the coupling strength for given photon number $n$. The Hamiltonian $H^{\left(n\right)}$ can be diagonalized for all choices of the physical parameters, see the Appendix \ref{sec:diag}. Here, we focus on the particular case with $\omega_{a}=\omega_c$ and obtain the atom-light dressed states
\begin{align}
 \left|D_{n}\right\rangle   &= N_n (2g_{n} \left|D\right\rangle \left|n\right\rangle  - \Delta\left|G\right\rangle \left|n+1\right\rangle)), \label{eq:Ddressed} \\
\left|B_{\pm,n}\right\rangle  & = -(N_n /\sqrt{2}) (\Delta\left|D\right\rangle \left|n\right\rangle + 2g_{n}\left|G\right\rangle \left|n+1\right\rangle) \nonumber \\
& \mp (1/\sqrt{2})\left|B\right\rangle \left|n\right\rangle, \label{eq:Bdressed}
\end{align}
with the frequencies $\omega_{0,n}=\left(n+1\right)\omega_{c},\  \omega_{\pm,n}=\left(n+1\right)\omega_{c}\pm\sqrt{g_{n}^{2}+\left(\Delta/2\right)^{2}}$  and the factor $N_n = [ \Delta^{2} + 4g_{n}^{2}) ]^{-1/2}$.

\emph{Magnetic Field-controlled Transmission} We first consider the three dressed states $\left|D_{0}\right\rangle, \left|B_{\pm,0}\right\rangle$ with lowest frequencies, $\omega_{0,0}= \omega_{c},\omega_{\pm,0}=\omega_{c}\pm\sqrt{2Ng^{2}+ \Delta^2/4}$, which are obtained from Eqs. (\ref{eq:Ddressed}) and (\ref{eq:Bdressed}) by seting $n=0$.  We see that the dressed states $\left|D_{0}\right\rangle,\left|B_{\pm,0}\right\rangle$  depend on the ground state $\left|G\right\rangle$ and the dark atomic state  $\left|D\right\rangle$, respectively, due to the magnetic field-induced Zeeman splitting $\Delta$, which can be understood as a mixing of the dark and bright excited states. In addtion, the excitation frequency of $\left|D_{0}\right\rangle$  coincides with the bare cavity frequency while that of $\left|B_{\pm,0}\right\rangle$ differ from the cavity mode, see Fig. \ref{fig:system} (d). Thus, these dressed states can be resonantly excited and their photonic components cause the three transmission peaks \citep{MNWinchester}, see Fig. \ref{fig:transmission} (a). Further more, the intensities of the center and side peaks are proportional to the populations $\Delta^2/(\Delta^2+ 8Ng^2), 8Ng^2/(\Delta^2+ 8Ng^2)$ of the higher photonic components in the dressed states. Thus, as the Zeeman-splitting  $\Delta$ increases, the center peak becomes stronger while the side peaks separate and become weaker, see Fig. \ref{fig:transmission} (a). This is precisely what is observed in the experiment  \citep{MNWinchester}. In addition, when the frequency $\omega_d$ of a probe laser is swept across the resonance frequencies, $\omega_{0,0}, \omega_{\pm,0}$,  the frequency difference between the probe laser and the dressed states changes sign and this leads to a phase change in the transmitted signal around the resonant frequencies as shown in Fig. \ref{fig:transmission} (b).

If we probe the system with a stronger laser, it is possible to excite the dressed states $\left|B_{\pm,n}\right\rangle$ and $  \left|D_{n}\right\rangle$ with higher photon number $n>0$, see Fig. \ref{fig:dressed-states} in the Appendix \ref{sec:diag}. As a result,  we expect more peaks in the transmission spectrum, see Fig. \ref{fig:transmission} (c,d), associated with the nine resonant transitions as indicated in Fig. \ref{fig:dressed-states} in the Appendix \ref{sec:diag}. These transitions are among the dark or bright dressed states (the first group), between the dark and bright dressed states (the second group), and between the bright states of different sign (the third group).   So far, we restrict the analysis to the states with at most a single atomic excitation but will relax this restriction in the following.

\emph{Master Equation} In the above analysis, we disregarded the dissipation in the system such as photon loss and spontaneous emission. To model these processes and also the atomic incoherent excitation, which is prerequisite for lasing, we establish a master equation (see the Appendix \ref{sec:smf}):
\begin{align}
 & \frac{\partial}{\partial t}\rho=-\frac{i}{\hbar}\left[H_{a}+H_{c}+H_{a-c}+H_{d},\rho\right]-\kappa\mathcal{D}\left[a\right]\rho \nonumber \\
 & -\sum_{k} \frac{1}{2}(\gamma_{+,k} - \gamma_{-,k}) \left(\mathcal{D}\left[A_{gD}^{k},A_{Bg}^{k}\right]\rho+\mathcal{D}\left[A_{Bg}^{k},A_{gD}^{k}\right]\rho\right)  \nonumber  \\
 & -\sum_{k} \frac{1}{2}(\gamma_{+,k} + \gamma_{-,k}) \left(\mathcal{D}\left[A_{gB}^{k}\right]\rho+\mathcal{D}\left[A_{gD}^{k}\right]\rho\right)  \nonumber  \\
 & -\sum_{k}\frac{1}{2}(\eta_{+,k} - \eta_{-,k})\left(\mathcal{D}\left[A_{Dg}^{k},A_{gB}^{k}\right]\rho+\mathcal{D}\left[A_{gB}^{k},A_{Dg}^{k}\right]\rho\right) \nonumber  \\
 & -\sum_{k}\frac{1}{2}(\eta_{+,k} + \eta_{-,k}) \left(\mathcal{D}\left[A_{Bg}^{k}\right]\rho+\mathcal{D}\left[A_{Dg}^{k}\right]\rho\right).   \label{eq:meq}
\end{align}
Here, $H_{d}=\sqrt{\kappa_{1}}\hbar\Omega (e^{-i\omega_{d}t}a^+ +\mathrm{h.c.})$ describes the coupling between the cavity mode and the probe laser with a strength $\Omega$ and a frequency  $\omega_d$, and $\sqrt{\kappa_{1}}$ is the transmission coefficient of the left mirror. The superoperators are defined as $\mathcal{D}\left[o\right]\rho=\left\{ o^{+}o,\rho\right\} /2-o\rho o^{+}$ and  $\mathcal{D}\left[o,p\right]\rho=\left\{ po,\rho\right\} /2-o\rho p$ (with any operator $o,p$), describing cavity loss with the rate $\kappa=\kappa_{1}+\kappa_{2}=2\pi \times150$ kHz due to the left ($\kappa_{1}$) and right mirror ($\kappa_{2}$), and atomic spontaneous emission with rates $\gamma=\gamma_{+,k}=\gamma_{-,k}=2\pi\times 7.5$ kHz. Eq.(\ref{eq:meq}) includes also the incoherent atomic excitation with rates $\eta_{+,k},\eta_{-,k}$, e.g., obtained via excitation of higher short-lived excited states, see Fig.1(b).  For simplicity, we have introduced the abbreviations $A^k_{st} = \left|s_{k}\right\rangle \left\langle t_{k}\right|$ ($s,t=g,B,D$) and ignored the negligible pure atomic dephasing in the optical lattice clock system.  Note that the dissipation introduces a dissipative coupling between the dark and bright atomic states as captured by terms of the form $\mathcal{D}\left[o,p\right]\rho$.

To simulate systems with thousands of atoms, we utilize second-order mean-field theory \cite{KDebnath}. In this theory, we derive the equation $\partial \left\langle o \right\rangle /\partial t = {\rm tr} \{o \partial \rho /\partial t \}$ with Eq. (\ref{eq:meq}) for the expectation value $\left\langle o \right\rangle$ of any observable $o$, see the Appendix \ref{sec:smf}. The equation for the mean photon number in the cavity $\left\langle a^{+}a\right\rangle $ couples to the atom-photon correlations $\left\langle aA^k_{st} \right\rangle $  which in turn depend on the atom-atom correlations $\langle A^k_{st}A^{k'}_{s't'} \rangle $ ($k \neq k'$) and third-order correlations, e.g. $\left\langle a^+ a A^k_{st} \right\rangle$. To truncate the hierarchy of equations, we approximate third-order quantities with products of lower-order terms, e.g. $\left\langle a^{+}aA^k_{st}\right\rangle =\left\langle a^{+}a\right\rangle \left\langle A^k_{st}\right\rangle +\left\langle a^+\right\rangle\left\langle aA^k_{st}\right\rangle +\left\langle a^+ A^k_{st}\right\rangle \left\langle a\right\rangle -2 \left\langle a^+\right\rangle  \left\langle a\right\rangle  \left\langle A^k_{st}\right\rangle$, resulting in closed non-linear equations, which involve these quantities and also the photon-photon correlation $\left\langle aa\right\rangle $,  the cavity field amplitude $\left\langle a \right\rangle $, the atomic state populations $\left\langle A^k_{ss} \right\rangle $ and the atomic polarization $\left\langle A^k_{st} \right\rangle $ ($s\neq t$). Through the atom-atom correlations, the equations capture atomic collective effects and lasing involving multiply excited atomic states. Assuming identical properties of all atoms, $\left\langle A^k_{st} \right\rangle $ and $\left\langle a A^k_{st} \right\rangle $ are the same for all $k$, and $\langle A^k_{st} A^{k'}_{s't'} \rangle $ are identical for all pairs ($k,k'$). This allows us to reduce the number of independent elements to just $102$ and their equations can be readily solved.

Fig. \ref{fig:transmission} shows the calculated transmission spectrum of a system with $6.25\times10^4$ atoms. Here, we illuminate the system by a Gaussian laser pulse $\Omega =\Omega_{0}\exp\left\{ -\left(t-\tau\right)^{2}/\left[2\sigma^{2}\right]\right\} $ with  maximum amplitude $\Omega_{0}$, peak time $\tau$ and duration $\sigma$, and evaluate the transmission as the ratio between the Fourier transform of the time-dependent input strength $\sqrt{\kappa_1}\Omega$ and the output amplitude $\sqrt{\kappa_{2}} \left |\left\langle a\right\rangle \right|$ (here and in the following, we assume $\kappa_{1}=\kappa_{2}$). The results agree with the experiment \citep{MNWinchester}, with the parameters $\sigma = 26.4$ ns, $\tau = 264.1$ ns, $\Omega_0 = 10 \sqrt{{\rm kHz}}$  (a,b) or $\Omega_0 = 400 \sqrt{{\rm kHz}}$ (c,d).

\begin{figure}[!ht]
\begin{centering}
\includegraphics[scale=0.29]{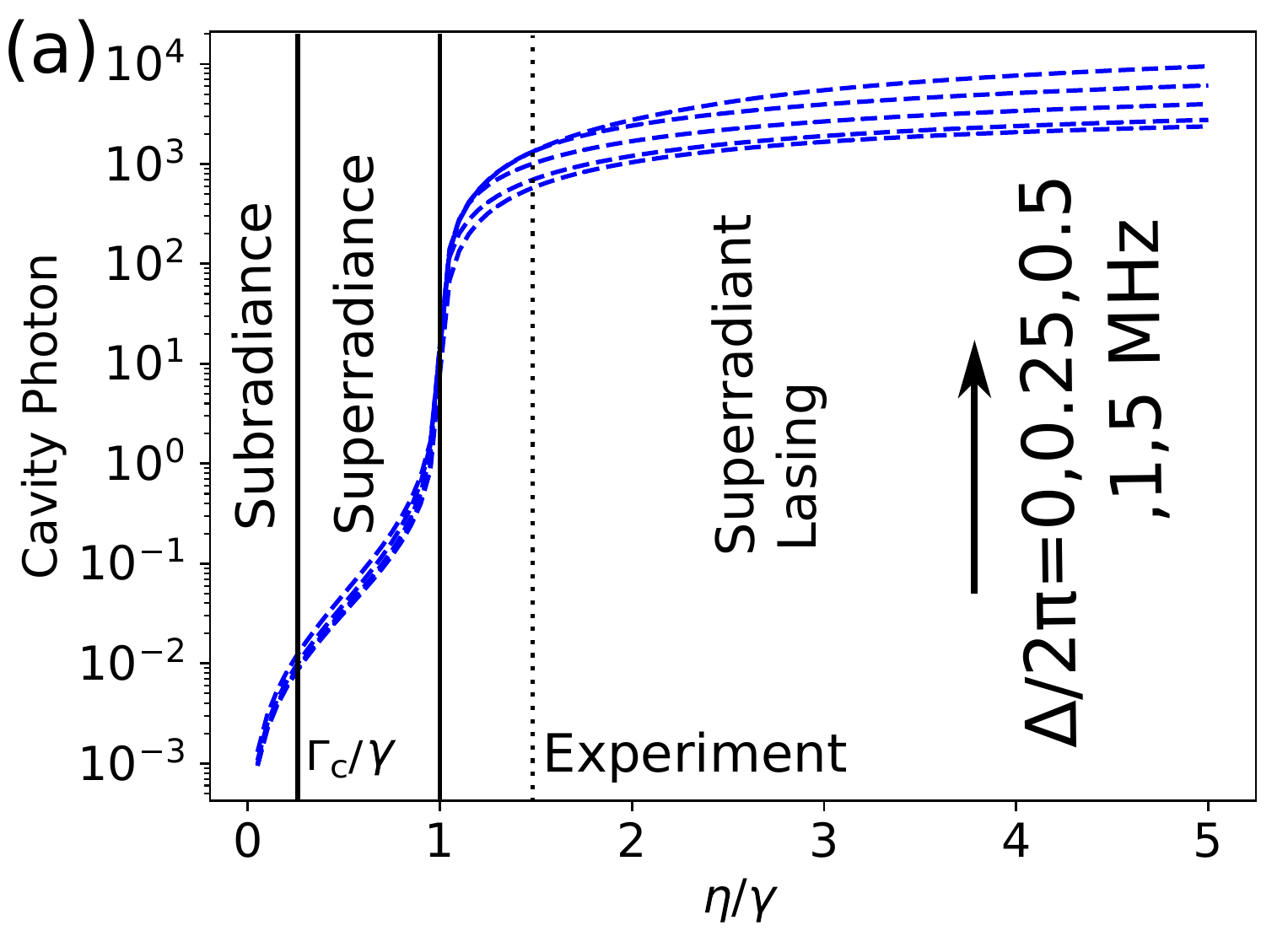}
\includegraphics[scale=0.29]{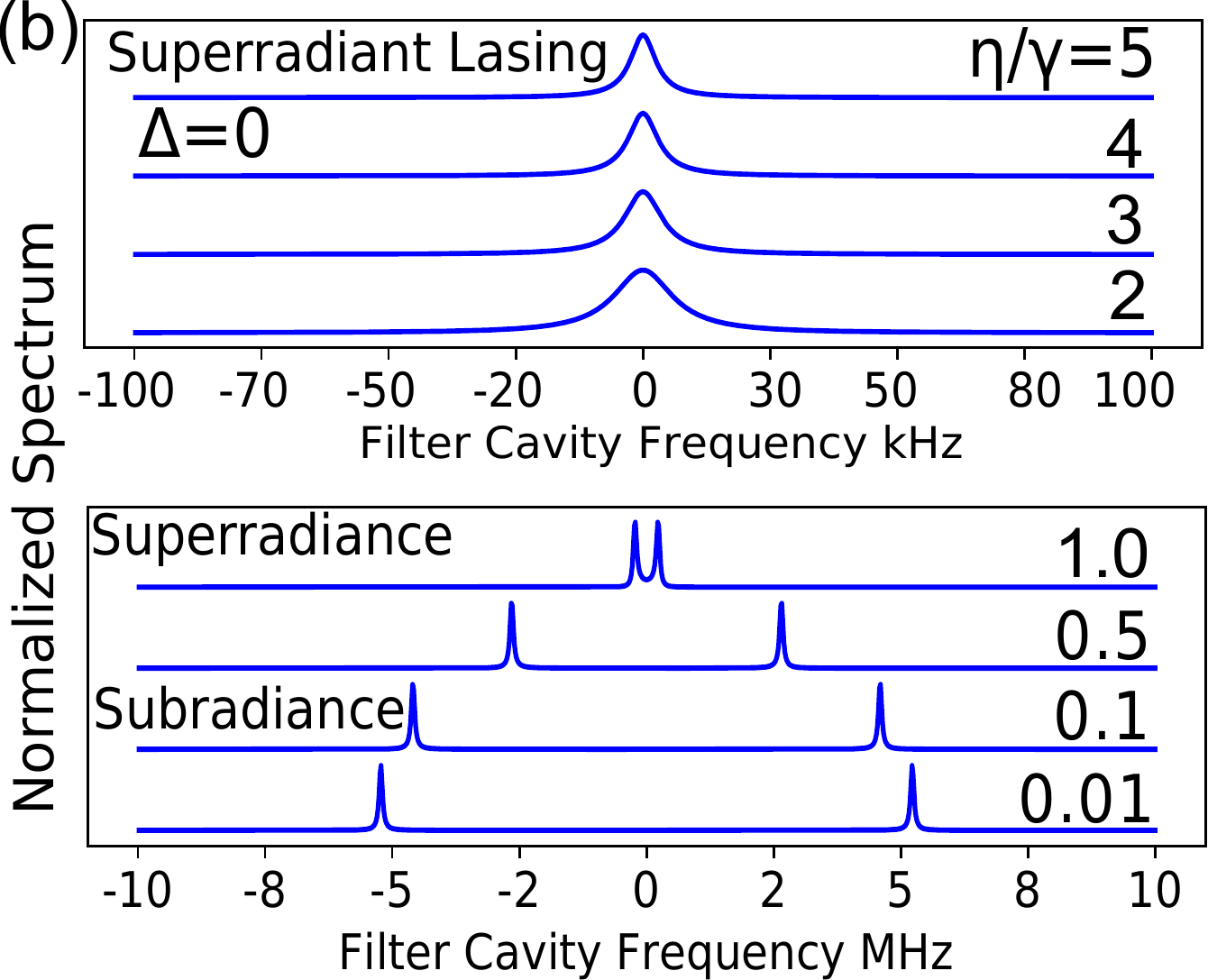}
\includegraphics[scale=0.31]{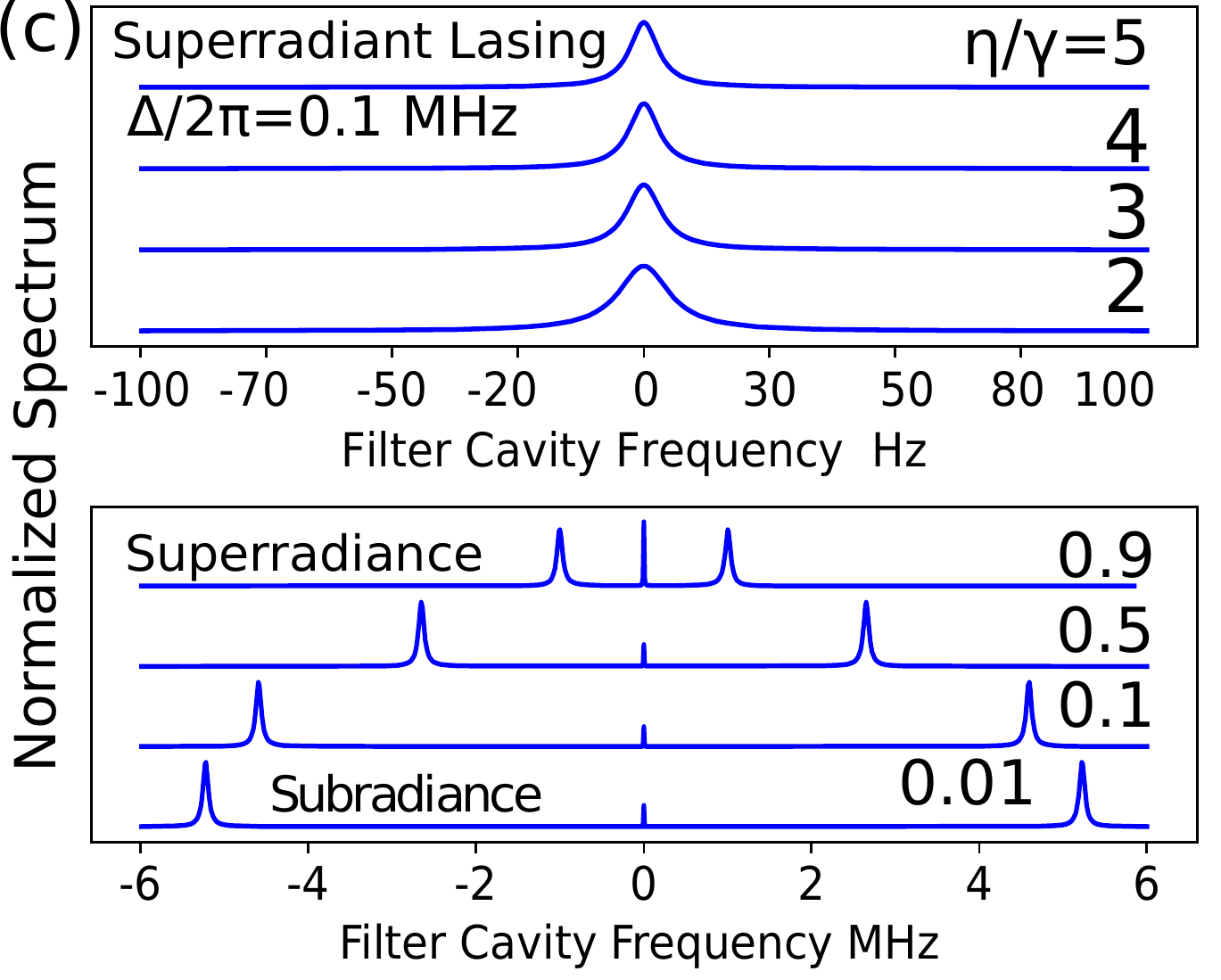}
\includegraphics[scale=0.33]{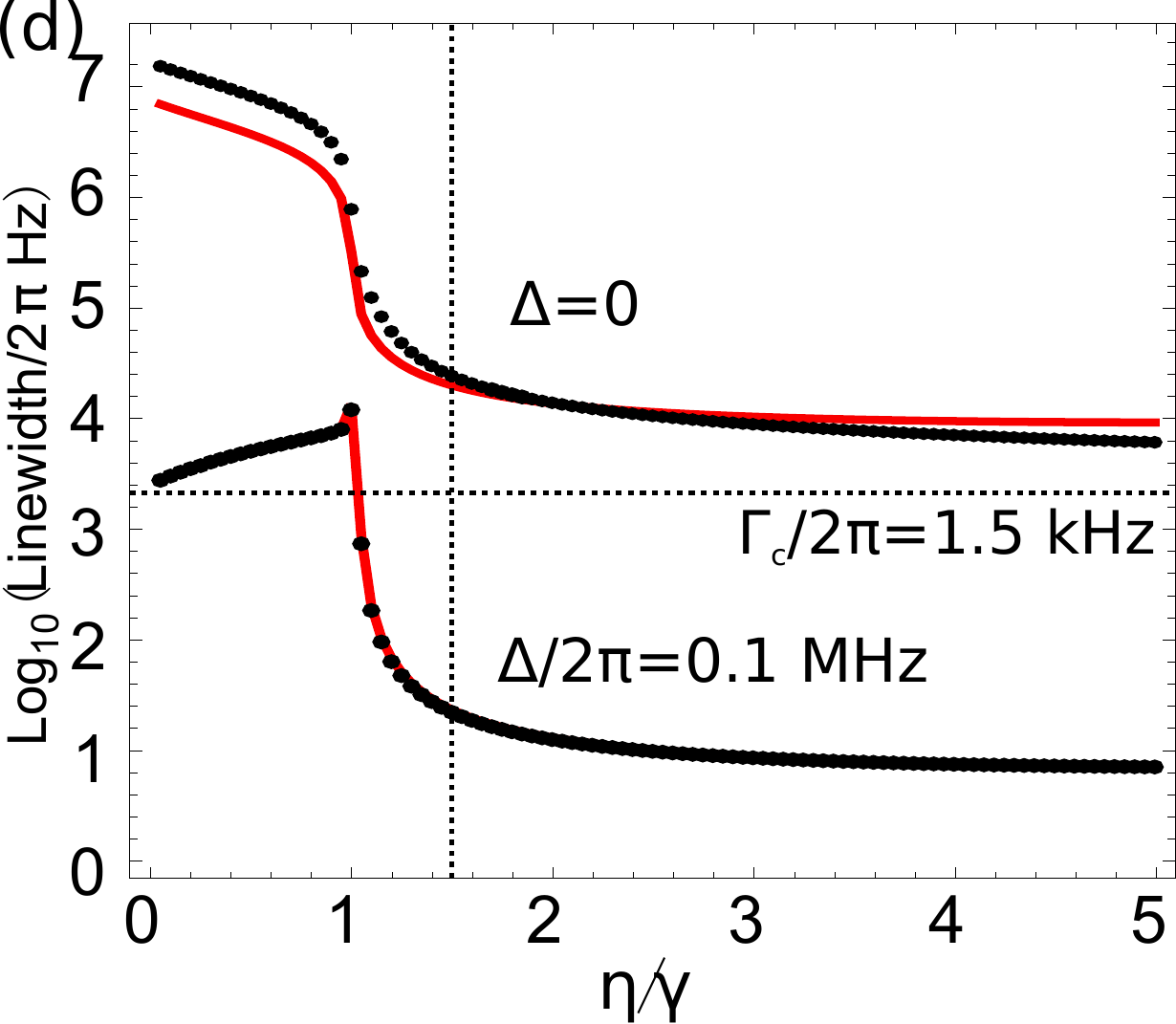}
\par\end{centering}
\caption{\label{fig:magnetic-lasing}
Magnetic field-controlled subradiance, superradiance and superradiant lasing in systems with $2.5\times10^5$ atoms. Panel (a) shows the intracavity photon number versus the incoherent pumping rate $\eta$ for increasing Zeeman splitting $\Delta$. Panels (b) and (c) show the normalized steady-state spectra for different $\eta$, without ($\Delta=0$) and with Zeeman splitting $\Delta/2\pi= 0.1$ MHz [note that in the upper panel (c) the frequency scale is in Hz]. Panel (d) compares the spectrum linewidth in the system without (upper curves) and with (lower curves) Zeeman splitting. The black dots and the red curves are the accurate results and those given by the semi-analytical expression (\ref{eq:analyticlinewidth}), respectively. The vertical and horizontal dotted line in (d) indicate the pumping reported in \cite{MANorcia} and the Purcell enhanced rate $\Gamma_c$, respectively. Other parameters are specified in the text. }
\end{figure}

\emph{Magnetic Field-controlled Superradiant Lasing} We now consider a cavity that is not driven by a laser but the atoms are incoherently pumped, e.g, through the coherent excitation of the short-lived higher excited state in Fig.1(b). We consider the continuous pumping $\eta=\eta_{+,k}=\eta_{-,k}$ and hence a continuous steady-state release of the atomic excitation energy via the cavity mode. In the absence of Zeeman-splitting $\Delta=0$, the three-level atoms behave like two-level systems since the dark atomic excited state is incoherently populated by the pump but couples with neither the bright atomic excited state nor the cavity mode.  As discussed in Ref. \citep{MXu,KDebnath}, this system undergoes transitions from subradiance to superradiance and finally to superradiant lasing as the pumping strength $\eta$ overcomes the Purcell enhanced decay rate $\Gamma_c=4g^2/\kappa = 2\pi\times1.5$ kHz and the atomic decay rate $\gamma$, respectively, see the lower curve in Fig. \ref{fig:magnetic-lasing} (a) for $\Delta=0$.

\begin{figure}[!ht]
\begin{centering}
\includegraphics[scale=0.33]{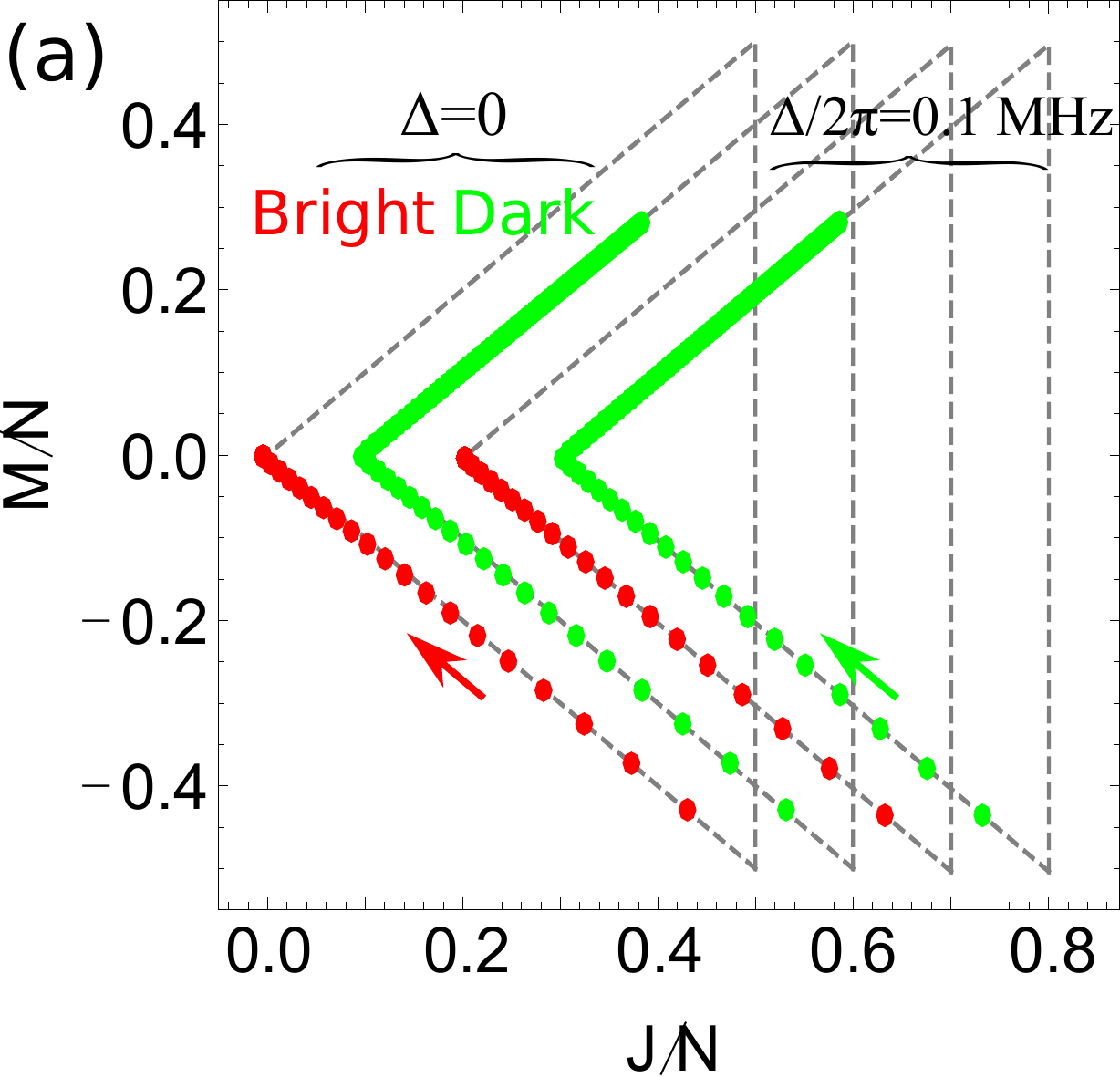}
\includegraphics[scale=0.33]{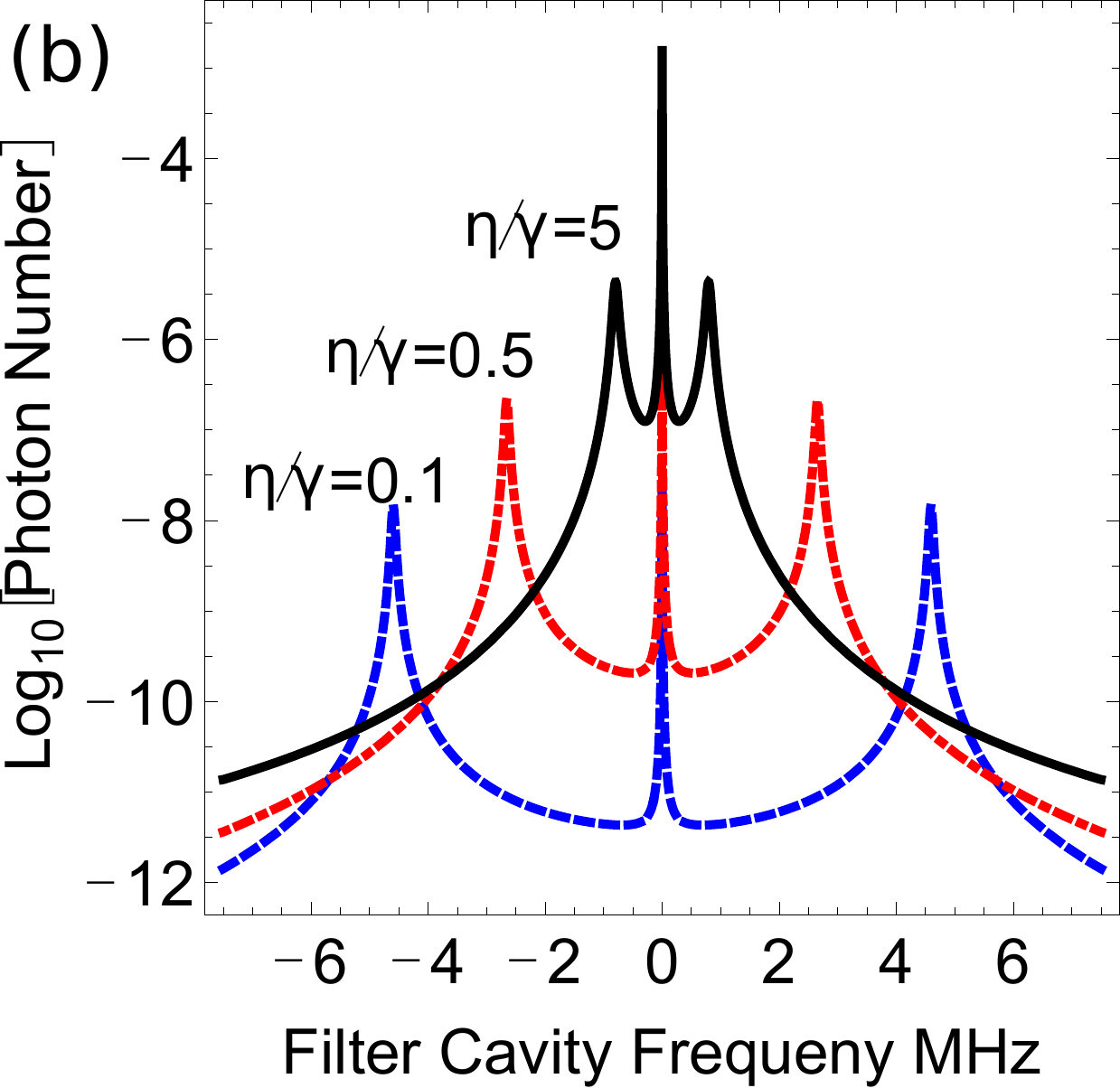}
\par\end{centering}
\centering{}\caption{\label{fig:DickePicture} (a) Evolution of the quantum numbers $J_s,M_s$ of pseudo Dicke states $\left| J_s,M_s \right\rangle$ associated with the bright $s=B$ (red dots) and dark $s=D$ (green dots) atomic transition as the pumping strength $\eta$ increases (arrows) for the systems without (left) and with (right) Zeeman-splitting $\Delta=2\pi\times0.1$ MHz. The results are shifted for the sake of clarity and the dashed lines indicate the boundaries of each set of Dicke ladders.  (b) Optical emission spectra on logarithmic scale for the pumping $\eta/\gamma = 0.1, 0.5, 5$ in Fig. \ref{fig:magnetic-lasing} (c). Other parameters are specified in the text.}
\end{figure}

To  calculate the spectrum of the emitted radiation, we mimic the spectral measurement by coupling the cavity output to a narrow-linewidth filter cavity, and then calculate the photon number in the filter cavity as a function of its frequency \citep{KDebnath}, see the Appendix \ref{sec:SpeFilterCavity}, \ref{sec:steady-state}.  Fig. \ref{fig:magnetic-lasing}(b) shows the spectrum for a system with $2.5\times 10^5$ atoms in the absence of Zeeman splitting $\Delta=0$. As the system is pumped more strongly it undergoes the lasing transition. The two peaks in the spectrum for weak pumping can be attributed to transitions between the bright dressed states $\left| B_{\pm,0} \right\rangle $ and the ground state in Fig. \ref{fig:system} (d). For stronger pumping, the peaks approach each other because the small incoherent pumping reduces the effective collective coupling $J \leq N/2$ of the atomic ensemble with the cavity mode \citep{KDebnath,YZhang1,NShammah}, which is shown in Fig. \ref{fig:DickePicture} (a). There, we visualize the collective atomic excitation by treating the bright $B_k\to g_k$ and dark $D_k\to g_k$ transitions as separate two-level transitions and introducing corresponding pseudo Dicke states \citep{RHDick}, see the Appendix \ref{sec:Dicke}. We see that as the incoherent pumping $\eta$ increases, the bright transition explores the lower rung of sub-radiant Dicke states due to the incoherent pumping but terminates at the states of lowest symmetry due to the balanced absorption and stimulated emission, while the dark state undergoes no stimulated emission process and hence it explores also states with $J\geq M > 0$, see the leftmost red and green dots in Fig. \ref{fig:DickePicture}(a), respectively.

In the presence of Zeeman splitting $\Delta/2\pi=0.1$ MHz, the dark atomic excited states couple with the bright states and thus influence the intra-cavity photon number and the spectrum. For $\eta>\gamma$, a finite value of $\Delta$ thus makes the intra-cavity photon number increase by about ten times, see Fig. \ref{fig:magnetic-lasing} (a). In addition, for weak and intermediate pumping $\eta$, we find a new peak in the spectrum around the cavity mode frequency, see Fig. \ref{fig:magnetic-lasing} (c). This peak can be attributed to transitions between the dark dressed state  $\left| D_{0} \right\rangle $ and the ground state in Fig. \ref{fig:system} (d). As $\eta$ increases, the side peaks become weaker and ultimately indiscernible while the center peak gets stronger and more narrow. For more results on the influence of the Zeeman-splitting on the spectrum, see Fig. \ref{fig:linewidth-DB}(a) in the Appendix \ref{sec:analytical-linewidth}. In this case, the Dicke state evolution is very similar, see  Fig. \ref{fig:DickePicture} (a), except that $M_B<0$ is negative (note $M_B>0$ is positive for $\Delta=0$ and larger $\eta$).

In Fig. \ref{fig:magnetic-lasing}(d), we compare the spectrum linewidth for the systems with (upper black dots) and without Zeeman-splitting $\Delta$ (lower black dots). For $\Delta=0$ and weak pumping $\eta<\gamma$ the linewidth corresponds to the frequency difference between the two side peaks. Remarkably, the linewidth for $\Delta\neq0$ decreases to about $10$ Hz for stronger pumping $\eta$, which is orders of magnitude smaller than the linewidth of $2\pi \times 10$ kHz for $\Delta=0$, the atomic decay rate of $2\pi \times 7.5 $ kHz, the pumping rate $2\pi \times 37.5$ kHz, the cavity loss rate of $2\pi \times 150$ kHz and the Purcell decay rate  $\Gamma_c= 2\pi \times 1.5 $ kHz.

To understand the ultra-narrow linewidth, we have derived a semi-analytical expression:
\begin{align}
& \Gamma  = \frac{\kappa  -\theta \bigl[ ( \gamma  + 2\eta)\left(\left\langle A_{BB}^k \right\rangle -\left\langle A_{gg}^k \right\rangle \right)  - \Delta{\rm Im}\left\langle A_{DB}^k \right\rangle \bigr] } {2+\theta (\left\langle A_{BB}^k \right\rangle-\left\langle A_{gg}^k \right\rangle)},  \label{eq:analyticlinewidth}
\end{align}
with the parameter $\theta = 8Ng^2/[\left( \gamma+ 2\eta\right)^2 + \Delta^2] $, see the Appendix \ref{sec:analytical-linewidth}. Fig. \ref{fig:magnetic-lasing} (d) shows that  Eq. \eqref{eq:analyticlinewidth}  (red curves) reproduces the accurate results (black dots) very well. Eq. \eqref{eq:analyticlinewidth} indicates that the linewidth depends only on the population inversion $\left\langle A_{BB}^k \right\rangle -\left\langle A_{gg}^k \right\rangle >0$ between the bright excited and the ground states in the absence of Zeeman-splitting $\Delta = 0$, while it depends crucially on the coherence $\left\langle A_{DB}^k \right\rangle$ between the bright and dark states for $\Delta\neq0$.  For the optimal pumping rate $\eta$ and Zeeman splitting $\Delta$, we observe an almost perfect cancellation of these contributions in the enumerator of Eq. \eqref{eq:analyticlinewidth}, leading to the ultra-narrow spectrum, see Fig. \ref{fig:linewidth-DB} (d) of the Appendix \ref{sec:analytical-linewidth}.

For weak  pumping, the spectrum is explained by the three dressed atom-light states formed by the first excited state of the atomic ensemble, see Fig. \ref{fig:system}(d), and it is consistent that for strong pumping, the ultra-narrow spectrum is due to the dressed atom-light states formed by the highly excited states of the atomic ensemble. Thus, in principle, we can construct a similar energy diagram as Fig. \ref{fig:dressed-states} of the Appendix \ref{sec:diag} with adjacent highly excited states of the atomic ensemble, and expect (1) coherence between the bright and dark atomic state due to the Zeeman-splitting induced state-mixing, and (2) continuation of the three spectral peaks for strong pumping. Indeed, by displaying the spectra in Fig.  \ref{fig:magnetic-lasing} (c) on a  logarithmic scale, we do observe two side peaks with extremely small intensity outside the ultra-narrow and intense peak, see Fig. \ref{fig:DickePicture} (b).

The lasing mechanism bears strong resemblance with the lasing without inversion mechanism (LWI)  \citep{JMompart}, such as the formation of dressed states and the absence of population inversion. Conventional LWI involves three-level systems interacting with a strong driving field, leading to coherence in the system and dressed states, and with a weak probe field, which is amplified due to  transitions among the dressed states. In our system, the quantized cavity field and the Zeeman coupling play roles analogous to the driving field and the probe field in LWI, the dark and bright dressed states are formed and evolve due to both couplings, and the incoherent pumping drives the population and coherence dynamics and hence the emission by the system, see the Appendix \ref{sec:lwi}.

\paragraph{Conclusion}
In summary, our theory explains the magnetic field-controlled optical transmission demonstrated in  Ref. \citep{MNWinchester} and predicts more transmission peaks for stronger probe fields associated with higher energy atom-light dressed states. With incoherently pumped atoms, we predict steady-state superradiant lasing with a remarkable orders of magnitude reduction of the laser linewidth in the presence of a magnetic field. A theory of lasing without inversion incorporating collective and superradiant effects  may provide further insight in our results. We anticipate that the super-narrow lasing may find applications in optical atomic clocks with strontium atom \citep{SASchaffer,CChen} and other atoms like calcium and ytterbium.

This work was supported by the Villum Foundation.

\newpage
\appendix

\renewcommand{\thefigure}{A\arabic{figure}}
\setcounter{figure}{0}

\section{Hamiltonian Diagonalization \label{sec:diag}}
In this section, we diagonalize the system Hamiltonian $H_s = H_{a} + H_{c} + H_{a-c}$  in the basis of dark and bright atomic states. To facilitate the derivation, we remind the reader about the atomic Hamiltonian
$H_{a}  =  \hbar \omega_{a}  \sum_{k=1}^{N} \left(\left|B_{k}\right\rangle \left\langle B_{k}\right|+\left|D_{k}\right\rangle \left\langle D_{k}\right|\right) +\hbar(\Delta/2)\sum_{k}\left(\left|B_{k}\right\rangle \left\langle D_{k}\right|+\left|D_{k}\right\rangle \left\langle B_{k}\right|\right)$ and the atom-cavity mode interaction $H_{a-c}  =  \hbar \sqrt{2}g \sum_k \left|B_{k}\right\rangle \left\langle g_{k}\right| a+h.c.$. Here, we assume that all the atoms are identical.

In the weak excitation limit, we can introduce the ground state  $\left|G\right\rangle =\prod_{k}\left|g_{k}\right\rangle $ and the bright and dark collective excited state $\left|B\right\rangle   = (1/\sqrt{N})\sum_{k}\left|B_{k}\right\rangle \prod_{j\neq k}\left|g_{k}\right\rangle$, $\left|D\right\rangle   =(1/\sqrt{N})\sum_{k}\left|D_{k}\right\rangle \prod_{j\neq k}\left|g_{k}\right\rangle$ of the atomic ensemble, and approximate the atomic ensemble Hamiltonian  as $H_{a} \approx \hbar\omega_{a}\left(\left|B\right\rangle \left\langle B\right|+\left|D\right\rangle \left\langle D\right|\right) +\hbar(\Delta/2)\left(\left|B\right\rangle \left\langle D\right|+\left|D\right\rangle \left\langle B\right|\right)$ and  the atom-cavity mode interaction Hamiltonian as  $H_{a-c} \approx\hbar\sqrt{2N}g\left|B\right\rangle \left\langle G\right|a+h.c.$.

\begin{figure}[!ht]
\begin{centering}
\includegraphics[scale=0.7]{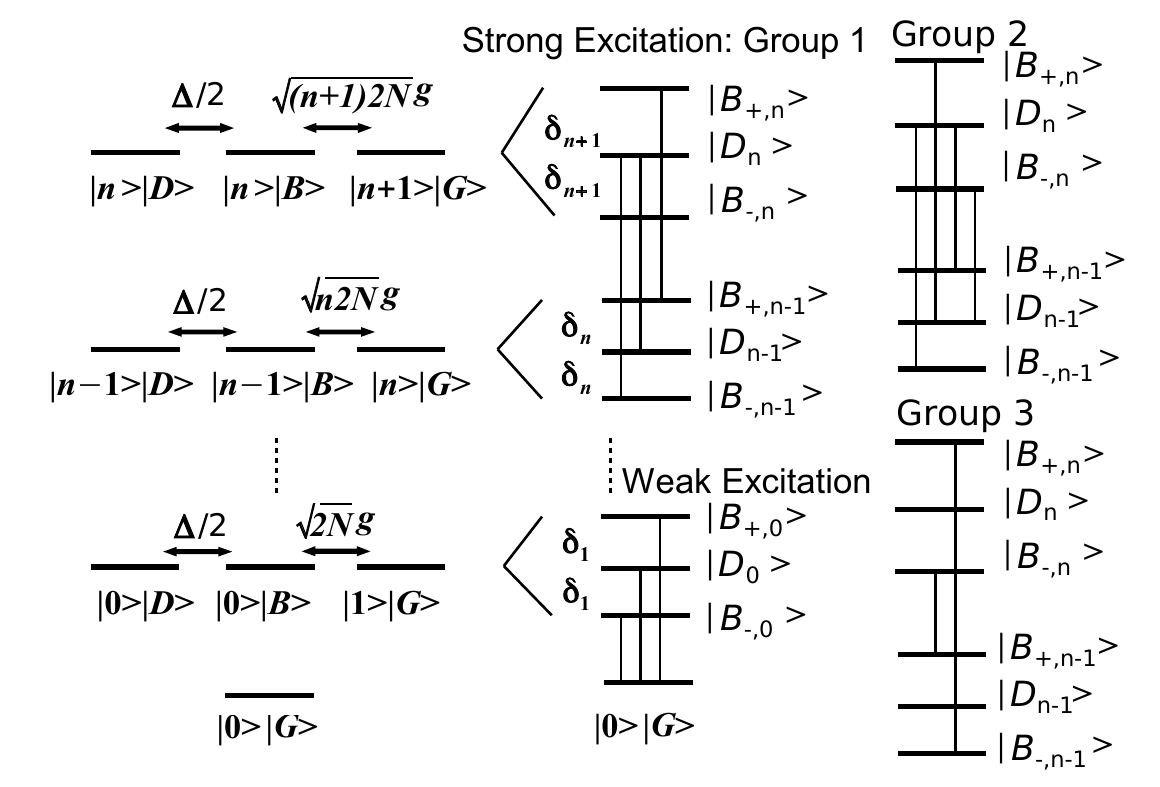}
\par\end{centering}
\caption{\label{fig:dressed-states} Energy diagram. The left part shows collective atom-photon product states and their interaction through the Zeeman-splitting $\Delta/2$ and the collective atom-cavity mode coupling $\sqrt{n2N}g$. The right part shows three groups of transitions (vertical lines) among the atom-light dressed states. }
\end{figure}

Following the spirit of the Jaynes-Cumming model for two-level atoms \cite{Jaynes}, we introduce the atom-photon product states  $\left|n\right\rangle \left|D\right\rangle $, $\left|n\right\rangle \left|B\right\rangle $, $\left|n+1\right\rangle \left|G\right\rangle $ and expand the approximated Hamiltonian $H_s=\oplus_n H^n$ in the basis of these states. For any given photon number $n$, $H^n$ is given by Eq. (1) in the main text. By diagonalizing this sub-Hamiltonian, we obtain three atom-light (photon) dressed states
\begin{align}
 \left|D_{n}\right\rangle   &= N_n [2\left(g_{n}^{2}+\omega_{ac}\delta_{0,n} -\delta_{0,n}^{2}\right)\left|D\right\rangle \left|n\right\rangle \nonumber \\
 & -\delta_{0,n}\Delta\left|B\right\rangle \left|n\right\rangle -g_{n}\Delta\left|G\right\rangle \left|n+1\right\rangle)], \\
\left|B_{\pm,n}\right\rangle  & = N_n [2\left(g_{n}^{2}+\omega_{ac}\delta_{\pm,n}-\delta_{\pm,n}^{2}\right)\left|D\right\rangle \left|n\right\rangle \nonumber \\
& -\delta_{\pm,n}\Delta\left|B\right\rangle \left|n\right\rangle -g_{n}\Delta\left|G\right\rangle \left|n+1\right\rangle],
\end{align}
with the frequencies
$\omega_{0,n}=\left(n+1\right)\omega_{c}+\delta_{0},\omega_{\pm,n}=\left(n+1\right)\omega_{c}+\delta_{\pm,n}$. Here, $\delta_{+,n},\delta_{0,n},\delta_{-,n}$ are the frequency shifts relative to the frequency   $\left(n+1\right)\omega_{c}$ of the photon number states $ \left|n+1\right\rangle$ and the normalization factors are defined as  $N_n = [4\left(g_{n}^{2}+\omega_{ac}\delta_{\pm,n}-\delta_{\pm,n}^{2}\right)^{2}+\left(\delta_{\pm,n}\Delta\right)^{2}+\left(g_{n}\Delta\right)^{2}]^{-1/2}$. Here, the label $+,0,-$ sorts the shifts in a descending order and we introduce the abbreviations $\omega_{ac}=\omega_a -\omega_c$ and $g_n=\sqrt{n+1}\sqrt{2N}g$. These shifts are the solutions of the equation
\begin{equation}
4g_{n}^{2}\omega_{ac}-4\left[g_{n}^{2}+\left(\Delta/2\right)^{2}-\omega_{ac}^{2}\right]\delta-8\omega_{ac}\delta^{2}+4\delta^{3}=0,
\end{equation}
and can be written as
\begin{align}
\delta_{+,n} & =\frac{1}{6\gamma}\left(\epsilon+4\omega_{ac}\gamma+\gamma^{2}\right),\\
\delta_{0,n} & =\frac{1}{12}\left(8\omega_{ec}-\left(1-\sqrt{3}i\right)\frac{\epsilon}{\gamma^{2}}-\left(1+\sqrt{3}i\right)\gamma\right),\\
\delta_{-,n} & =\frac{1}{12}\left(8\omega_{ec}-\left(1-\sqrt{3}i\right)\gamma-\left(1+\sqrt{3}i\right)\frac{\epsilon}{\gamma^{2}}\right),
\end{align}
with the abbreviations $\epsilon=12g_{n}^{2}+3\Delta^{2}+4\omega_{ac}^{2}$,
$\gamma=\sqrt[3]{\alpha+3\sqrt{-3\left(4g_{n}^{2}+\Delta^{2}\right)^{3}+24\beta\omega_{ac}^{2}-48\Delta^{2}\omega_{ac}^{4}}}$,
$\beta=-2g_{n}^{4}-10g_{n}^{2}\Delta^{2}+\Delta^{4}$, $\alpha=-2\omega_{ac}\left(18g_{n}^{2}-9\Delta^{2}+4\omega_{ac}^{2}\right)$.  Although there are complex numbers in the above expressions, the  shifts are always real.  For the particular case with $\omega_c = \omega_a$, the above results are reduced to Eqs. (2) and (3) in the main text. In Fig. \ref{fig:dressed-states}, we show the formation of the dressed states and possible transitions between them.

\section{Second-order Mean-field Equations \label{sec:smf}}
In this section, we show how to solve the master equation with second-order mean-field theory. Firstly, we recapture the master equation
\begin{align}
 & \frac{\partial}{\partial t}\rho=-\frac{i}{\hbar}\left[H_{a}+H_{c}+H_{a-c} + H_{d},\rho\right]-\kappa\mathcal{D}\left[a\right]\rho\nonumber \\
 & -\sum_{k} \gamma_{+,k} \mathcal{D}\left[ A^k_{g+} \right]\rho-\sum_{k}\gamma_{-,k} \mathcal{D}\left[  A^k_{g-}  \right]\rho\nonumber \\
 & -\sum_{k} \eta_{+,k}\mathcal{D}\left[ A^k_{+g} \right]\rho-\sum_{k} \eta_{-,k} \mathcal{D}\left[A^k_{-g} \right]\rho. 
\end{align}
The Hamiltonians $H_{a},H_{c},H_{a-c},H_{d}$ have been introduced in the main text. The Lindblad terms describe the cavity photon loss with the rate $\kappa$,  the spontaneous emission of the atomic excited states $\left|e_{\pm,k}\right\rangle $ with the rates $\gamma_{\pm,k}$ and the pumping of these states with the rates  $\eta_{\pm,k}$. The superoperator is defined as $\mathcal{D}\left[o\right]\rho=\left\{ o^{+}o,\rho\right\} /2-o\rho o^{+}$ (with any operator $o$). For simplicity, we introduce the abbreviations $A^k_{st} = \left|s_{k}\right\rangle \left\langle t_{k}\right|$.

Using the expansion $ \left|e_{+,k}\right\rangle  =(1/\sqrt{2})\left(\left|B_{k}\right\rangle +\left|D_{k}\right\rangle \right)$, $\left|e_{-,k}\right\rangle  =(1/\sqrt{2})\left(\left|B_{k}\right\rangle -\left|D_{k}\right\rangle \right)$, we can transform the master equation (\ref{eq:meq}) to the equation (4) in the main text. For simplicity, we introduce the abbreviations   $\Gamma_{\pm,k} =\frac{1}{2}\left(\gamma_{+,k}\pm\gamma_{-,k}\right) $, $\Lambda_{\pm,k} =\frac{1}{2}\left(\text{\ensuremath{\eta}}_{+,k}\pm\text{\ensuremath{\eta}}_{-,k}\right) $.

From the master equation (4) in the main text, we can derive the equation $\partial \left\langle o \right\rangle /\partial t = {\rm tr} \{o \partial \rho /\partial t \}$ for the expectation value $\left\langle o \right\rangle$ of any observable $o$. Following this, we derive the equation for the photon number
\begin{align}
 & \frac{\partial}{\partial t}\left\langle a^{+}a\right\rangle =-\kappa\left\langle a^{+}a\right\rangle -\sqrt{\kappa_{1}}\Omega2\mathrm{Im}e^{i\omega_{d}t}\left\langle a\right\rangle \nonumber \\
 & -\sum_{k}2\sqrt{2}g_{k}\mathrm{Im}\left\langle aA_{Bg}^{k}\right\rangle,
\end{align}
which depends on the cavity mode amplitude $\left\langle a\right\rangle$ and the atom-photon correlation $\left\langle aA_{Bg}^{k}\right\rangle$. In the same way, we derive the equation
\begin{align}
 & \frac{\partial}{\partial t}\left\langle a\right\rangle =-\left(i\omega_{c}+\kappa/2\right)\left\langle a\right\rangle \nonumber \\
 & -i\sqrt{\kappa_{1}}\Omega e^{-i\omega_{d}t}-i\sum_{k}\sqrt{2}g_{k}\left\langle A_{gB}^{k}\right\rangle,
\end{align}
which depends on the polarization of the atoms $\left\langle A_{gB}^{k}\right\rangle$. In general, the equation for  $\left\langle A_{st}^{k}\right\rangle$ reads
\begin{align}
 & \frac{\partial}{\partial t}\left\langle A_{st}^{k}\right\rangle =-i\omega_{a}^{k}\sum_{r=B,D}\left(\delta_{t,r}\left\langle A_{sr}^{k}\right\rangle -\delta_{s,r}\left\langle A_{rt}^{k}\right\rangle \right)\nonumber \\
 & -i\left(\Delta_{k}/2\right)\sum_{r}\left(\delta_{t,r}\left\langle A_{s\bar{r}}^{k}\right\rangle -\delta_{s,\bar{r}}\left\langle A_{rt}^{k}\right\rangle \right)\nonumber \\
 & -i\sqrt{2}g_{k}\Bigl(\delta_{t,B}\left\langle aA_{sg}^{k}\right\rangle -\delta_{s,g}\left\langle aA_{Bt}^{k}\right\rangle \nonumber \\
 & +\delta_{t,g}\left\langle a^{+}A_{sB}^{k}\right\rangle -\delta_{s,B}\left\langle a^{+}A_{gt}^{k}\right\rangle \Bigr)\nonumber \\
 & -\Lambda_{+,k}\Bigl(\delta_{s,g}\left\langle A_{gt}^{k}\right\rangle +\delta_{t,g}\left\langle A_{sg}^{k}\right\rangle -\sum_{r}\delta_{s,r}\delta_{t,r}\left\langle A_{gg}^{k}\right\rangle \Bigr)\nonumber \\
 & +\Lambda_{-,k}\sum_{r}\delta_{s,r}\delta_{t,\bar{r}}\left\langle A_{gg}^{k}\right\rangle \nonumber \\
 & -\Gamma_{+,k}\sum_{r}\Bigl[\frac{1}{2}\left(\delta_{s,r}\left\langle A_{rt}^{k}\right\rangle +\delta_{t,r}\left\langle A_{sr}^{k}\right\rangle \right)-\delta_{s,g}\delta_{t,g}\left\langle A_{rr}^{k}\right\rangle \Bigr]\nonumber \\
 & -\Gamma_{-,k}\sum_{r}\Bigl[\frac{1}{2}\left(\delta_{s\bar{r}}\left\langle A_{rt}^{k}\right\rangle +\delta_{t,r}\left\langle A_{s\bar{r}}^{k}\right\rangle \right)-\delta_{s,g}\delta_{t,g}\left\langle A_{r\bar{r}}^{k}\right\rangle \Bigr].
\end{align}
In the above equation, $\bar{r}$ denotes $B,D$ when $r$ labels $D,B$, respectively. In addition, we also need the equation for the atom-photon correlations
\begin{align}
 & \frac{\partial}{\partial t}\left\langle aA_{st}^{k}\right\rangle =-\left(i\omega_{c}+\frac{1}{2}\kappa\right)\left\langle aA_{st}^{k}\right\rangle -i\sqrt{\kappa_{1}}\Omega e^{-i\omega_{d}t}\left\langle A_{st}^{k}\right\rangle \nonumber \\
 & -i\omega_{a}^{k}\sum_{r=B,D}\left(\delta_{t,r}\left\langle aA_{sr}^{k}\right\rangle -\delta_{s,r}\left\langle aA_{rt}^{k}\right\rangle \right)\nonumber \\
 & -i\left(\Delta_{k}/2\right)\sum_{r}\left(\delta_{t,r}\left\langle aA_{s\bar{r}}^{k}\right\rangle -\delta_{s,\bar{r}}\left\langle aA_{rt}^{k}\right\rangle \right)\nonumber \\
 & -i\sqrt{2}g_{k}\Bigl[\delta_{t,B}\left\langle aaA_{sg}^{k}\right\rangle -\delta_{s,g}\left\langle aaA_{Bt}^{k}\right\rangle -\delta_{s,B}\left\langle a^{+}aA_{gt}^{k}\right\rangle \nonumber \\
 & +\delta_{t,g}\left(\left\langle a^{+}aA_{sB}^{k}\right\rangle +\left\langle A_{sB}^{k}\right\rangle \right)\Bigr]-i\sum_{k'\neq k}\sqrt{2}g_{k'}\left\langle A_{gB}^{k'}A_{st}^{k}\right\rangle \nonumber \\
 & -\Lambda_{+,k}\Bigl(\delta_{s,g}\left\langle aA_{gt}^{k}\right\rangle +\delta_{t,g}\left\langle aA_{sg}^{k}\right\rangle -\sum_{r}\delta_{s,r}\delta_{t,r}\left\langle aA_{gg}^{k}\right\rangle \Bigr)\nonumber \\
 & +\Lambda_{-,k}\sum_{r}\delta_{s,r}\delta_{t,\bar{r}}\left\langle aA_{gg}^{k}\right\rangle \nonumber \\
 & -\Gamma_{+,k}\sum_{r}\Bigl[\frac{1}{2}\left(\delta_{s,r}\left\langle aA_{rt}^{k}\right\rangle +\delta_{t,r}\left\langle aA_{sr}^{k}\right\rangle \right)-\delta_{s,g}\delta_{t,g}\left\langle aA_{rr}^{k}\right\rangle \Bigr]\nonumber \\
 & -\Gamma_{-,k}\sum_{r}\Bigl[\frac{1}{2}\left(\delta_{s\bar{r}}\left\langle aA_{rt}^{k}\right\rangle +\delta_{t,r}\left\langle aA_{s\bar{r}}^{k}\right\rangle \right)-\delta_{s,g}\delta_{t,g}\left\langle aA_{r\bar{r}}^{k}\right\rangle \Bigr].
\end{align}

In the above equations, we encounter the expectation values of three operators, e.g. $\left\langle aaA_{Bt}^{k}\right\rangle$. If we derive the equations for these quantities, we will encounter the expectation values of four operators and so on, which creates a hierarchy of equations. To truncate this hierarchy, we apply the third order cumulant expansion  to approximate the expectation values of three operators, e.g.  $\left\langle aaA_{Bt}^{k}\right\rangle= \left\langle a \right\rangle\left\langle aA_{Bt}^{k}\right\rangle +\left\langle a\right\rangle \left\langle aA_{Bt}^{k}\right\rangle +  \left\langle A_{Bt}^{k}\right\rangle \left\langle aa\right\rangle - 2 \left\langle a\right\rangle \left\langle a\right\rangle \left\langle A_{Bt}^{k}\right\rangle$.  By doing so, we also need the equation for the photon-photon correlation  $\left\langle aa\right\rangle$:
\begin{align}
 & \frac{\partial}{\partial t}\left\langle aa\right\rangle =-\left(i2\omega_{c}+\kappa\right)\left\langle aa\right\rangle \nonumber \\
 & -i2\sqrt{\kappa_{1}}\Omega e^{-i\omega_{d}t}\left\langle a\right\rangle -i\sum_{k}2\sqrt{2}g_{k}\left\langle aA_{gB}^{k}\right\rangle .
\end{align}

Eq.(11) also depends on the atom-atom correlations $\left\langle A_{st}^{k}A_{s't'}^{k}\right\rangle $. The equation for these correlations is given by the rather lengthy Eq. (\ref{eq:atom-atom-correlation}).

\begin{figure*}[!ht]
\begin{align}
 & \frac{\partial}{\partial t}\left\langle A_{st}^{k}A_{s't'}^{k'}\right\rangle \nonumber \\
 & =-i\omega_{a}^{k}\sum_{r}\left(\delta_{t,r}\left\langle A_{sr}^{k}A_{s't'}^{k'}\right\rangle -\delta_{s,r}\left\langle A_{rt}^{k}A_{s't'}^{k'}\right\rangle \right)-i\left(\Delta_{k}/2\right)\sum_{r}\left(\delta_{t,r}\left\langle A_{s\bar{r}}^{k}A_{s't'}^{k'}\right\rangle -\delta_{s,\bar{r}}\left\langle A_{rt}^{k}A_{s't'}^{k'}\right\rangle \right)\nonumber \\
 & -i\sqrt{2}g_{k}\Bigl(\delta_{tB}\left\langle aA_{sg}^{k}A_{s't'}^{k'}\right\rangle -\delta_{sg}\left\langle aA_{Bt}^{k}A_{s't'}^{k'}\right\rangle +\delta_{t,g}\left\langle a^{+}A_{sB}^{k}A_{s't'}^{k'}\right\rangle -\delta_{s,B}\left\langle a^{+}A_{gt}^{k}A_{s't'}^{k'}\right\rangle \Bigr)\nonumber \\
 & -\Lambda_{+,k}\Bigl(\delta_{s,g}\left\langle A_{gt}^{k}A_{s't'}^{k'}\right\rangle +\delta_{t,g}\left\langle A_{sg}^{k}A_{s't'}^{k'}\right\rangle -\sum_{r}\delta_{s,r}\delta_{t,r}\left\langle A_{gg}^{k}A_{s't'}^{k'}\right\rangle \Bigr)+\Lambda_{-,k}\left(\delta_{s,B}\delta_{t,D}+\delta_{s,D}\delta_{tB}\right)\left\langle A_{gg}^{k}A_{s't'}^{k'}\right\rangle \nonumber \\
 & -\Gamma_{+,k}\sum_{r}\left[\frac{1}{2}\Bigl(\delta_{s,r}\left\langle A_{rt}^{k}A_{s't'}^{k'}\right\rangle +\delta_{t,r}\left\langle A_{sr}^{k}A_{s't'}^{k'}\right\rangle \Bigr)-\delta_{s,g}\delta_{t,g}\left\langle A_{rr}^{k}A_{s't'}^{k'}\right\rangle \right]\nonumber \\
 & -\Gamma_{-,k}\sum_{r}\left[\frac{1}{2}\Bigl(\delta_{s,\bar{r}}\left\langle A_{rt}^{k}A_{s't'}^{k'}\right\rangle +\delta_{t,r}\left\langle A_{s\bar{r}}^{k}A_{s't'}^{k'}\right\rangle \Bigr)-\delta_{s,g}\delta_{t,g}\left\langle A_{r\bar{r}}^{k}A_{s't'}^{k'}\right\rangle \right]\nonumber \\
 & -i\omega_{a}^{k'}\sum_{r}\left(\delta_{t',r}\left\langle A_{st}^{k}A_{s't'}^{k'}\right\rangle -\delta_{s',r}\left\langle A_{st}^{k}A_{s't'}^{k'}\right\rangle \right)-i\left(\Delta_{k'}/2\right)\sum_{r}\left(\delta_{t',r}\left\langle A_{st}^{k}A_{s'\bar{r}}^{k'}\right\rangle -\delta_{s',\bar{r}}\left\langle A_{st}^{k}A_{rt'}^{k'}\right\rangle \right)\nonumber \\
 & -i\sqrt{2}g_{k'}\Bigl(\delta_{t',B}\left\langle aA_{st}^{k}A_{s'g}^{k'}\right\rangle -\delta_{s',g}\left\langle aA_{st}^{k}A_{Bt'}^{k'}\right\rangle +\delta_{t',g}\left\langle a^{+}A_{st}^{k}A_{s'B}^{k'}\right\rangle -\delta_{s',B}\left\langle a^{+}A_{st}^{k}A_{gt'}^{k'}\right\rangle \Bigr)\nonumber \\
 & -\Lambda_{+,k'}\Bigl(\delta_{s',g}\left\langle A_{st}^{k}A_{gt'}^{k'}\right\rangle +\delta_{t',g}\left\langle A_{st}^{k}A_{s'g}^{k'}\right\rangle -\sum_{r}\delta_{s',r}\delta_{t',r}\left\langle A_{st}^{k}A_{gg}^{k'}\right\rangle \Bigr)+\Lambda_{-,k'}\left(\delta_{s',B}\delta_{t',D}+\delta_{s',D}\delta_{t'B}\right)\left\langle A_{st}^{k}A_{gg}^{k'}\right\rangle \nonumber \\
 & -\Gamma_{+,k'}\sum_{r}\left[\frac{1}{2}\Bigl(\delta_{s',r}\left\langle A_{st}^{k}A_{rt'}^{k'}\right\rangle +\delta_{t'',r}\left\langle A_{st}^{k}A_{s'r}^{k'}\right\rangle \Bigr)-\delta_{s',g}\delta_{t',g}\left\langle A_{st}^{k}A_{rr}^{k'}\right\rangle \right]\nonumber \\
 & -\Gamma_{-,k'}\sum_{r}\left[\frac{1}{2}\Bigl(\delta_{s',\bar{r}}\left\langle A_{st}^{k}A_{rt'}^{k'}\right\rangle +\delta_{t',r}\left\langle A_{st}^{k}A_{s'\bar{r}}^{k'}\right\rangle \Bigr)-\delta_{s',g}\delta_{t',g}\left\langle A_{st}^{k}A_{r\bar{r}}^{k'}\right\rangle \right]. \label{eq:atom-atom-correlation}
\end{align}
\end{figure*}

As we assume identical atoms , $\left\langle A^k_{st} \right\rangle $, $\left\langle a A^k_{st} \right\rangle $ are identical for any atom $k$ and $\langle A^k_{st} A^{k'}_{s't'} \rangle $ are identical for any atom pair $k,k'$. As a result, we can reduce the number of independent variables to $9(1+1)+9\times9+3=102$, where sums $\sum_k,\sum_{k'\neq k}$ are replaced with single terms multiplied by $N$ and $N(N-1)$ in the equations.

\section{Spectrum Computation with a Filter Cavity \label{sec:SpeFilterCavity}}

To calculate the lasing spectrum, we introduce a filter cavity and couple it to the main system by supplementing the master equation (4) in the main text with the terms $\left(\frac{\partial}{\partial t}\rho\right)_{m}=-\left(i/\hbar\right)\left[H_{f}+H_{f-c},\rho\right]-\chi\mathcal{D}\left[b\right]\rho$ \cite{KDebnath}.  The filter cavity Hamiltonian $H_{f}=\hbar\omega_{f}b^{+}b$ is specified by a frequency $\omega_{f}$, the creation $b^{+}$ and annihilation operator $b$ of photons. The filter cavity-system interaction $H_{f-c}=\hbar\beta\left(b^{+}a+a^{+}b\right)$ is specified with the coupling strength $\beta$, and the Lindblad term describes photon loss in the filter cavity with a rate $\chi$.

To calculate the spectrum, we consider the equation for the mean photon number $\left\langle b^{+}b\right\rangle $ in the filter cavity
\begin{equation}
\frac{\partial}{\partial t}\left\langle b^{+}b\right\rangle =-\chi\left\langle b^{+}b\right\rangle +\beta2\mathrm{Im}\left\langle b^{+}a\right\rangle , \label{eq:photon-filter}
\end{equation}
and  the field amplitude $\left\langle b\right\rangle $  in the filter cavity
\begin{equation}
\frac{\partial}{\partial t}\left\langle b\right\rangle =-\left(i\omega_{f}+\chi/2\right)\left\langle b\right\rangle -i\beta\left\langle a\right\rangle. \label{eq:amplitude-filter}
\end{equation}
The equation (\ref{eq:photon-filter})  depends on the photon-photon correlation between the main and filter cavity $\left\langle b^{+}a\right\rangle$, which follows the equation
\begin{align}
 & \frac{\partial}{\partial t}\left\langle b^{+}a\right\rangle =\left[i\left(\omega_{f}-\omega_{c}\right)-\left(\chi+\kappa\right)/2\right]\left\langle b^{+}a\right\rangle \nonumber \\
 & -i\beta\left(\left\langle b^{+}b\right\rangle -\left\langle a^{+}a\right\rangle \right)-i\sum_{k}\sqrt{2}g_{k}\left\langle b^{+}A_{gB}^{k}\right\rangle.
\end{align}
This equation depends on the photon-atom correlation $\left\langle b^{+}A_{gB}^{k}\right\rangle$, which follows the equation
\begin{align}
 & \frac{\partial}{\partial t}\left\langle b^{+}A_{gr}^{k}\right\rangle = -i\sqrt{2}g_{k}\left( \delta_{r,B} \left\langle b^{+}aA_{gg}^{k}\right\rangle -\left\langle b^{+}aA_{Br}^{k}\right\rangle \right) \nonumber \\
 & + \left[i\left(\omega_{f}-\omega_{a}^{k}\right)-\chi/2-\Lambda_{+,k}-\Gamma_{+,k}/2\right]\left\langle b^{+}A_{gr}^{k}\right\rangle \nonumber \\
 & +i\beta\left\langle a^{+}A_{gr}^{k}\right\rangle - \left(i\Delta_{k}/2+\Gamma_{-,k}/2\right) \left\langle b^{+}A_{g\bar{r}}^{k}\right\rangle.
\end{align}
In the above equation, $r=B,D$ indicates the quantities related to the bright and dark atomic excited state.  These quantities couple with each other through the Zeeman splitting $\Delta_{k}$ and through the incoherent pumping $\Lambda_{-,k} =\left(  \eta_{+,k} - \eta_{-,k} \right)/2$. Notice that the latter term exists only if the incoherent pumping $ \eta_{\pm,k}$ are different for the two extreme excited states $\left|e_{\pm,k}\right\rangle$.  In addition, the above equation depends on the expectation values of three operators, e.g. $\left\langle b^{+}aA_{gg}^{k}\right\rangle$, and we apply the third order cumulant expansion to approximate these values with low-order quantities.

To reduce the backaction of the filter cavity on the main system, we shall assume a small value for $\beta$. We shall also assume that $\chi$ is smaller than the linewidth of the spectrum that we want to measure.  Under these assumptions, we can first determine the steady-state expectation values of the system observables, and then obtain the filter cavity correlations and mean values for different filter-cavity frequencies $\omega_{f}$.

\section{ Systems without Atomic and Field Coherence \label{sec:steady-state} }
In the previous sections, we outlined the equations for rather general systems. However, if there is no initial coherence in the field or the atoms and the system is also not driven by the probe laser, the cavity field amplitude $\left\langle a\right\rangle$,$\left\langle b\right\rangle$ and the polarization $\left\langle A^k _{gr} \right\rangle$ vanish at all times. In this case, we can neglect all these quantities in all the equations to get the following small set of simplified equations.

The equation for the photon number can be now written as
\begin{align}
 & \frac{\partial}{\partial t}\left\langle a^{+}a\right\rangle =-\kappa\left\langle a^{+}a\right\rangle -\sum_{k}2\sqrt{2}g_{k}\mathrm{Im}\left\langle aA_{Bg}^{k}\right\rangle .
\end{align}
The equation for the atom-photon correlation becomes
\begin{align}
 & \frac{\partial}{\partial t}\left\langle aA_{rg}^{k}\right\rangle =\left[i\left(\omega_{a}^{k}-\omega_{c}\right)-\kappa/2-\Lambda_{+,k}-\Gamma_{+,k}/2\right]\left\langle aA_{rg}^{k}\right\rangle \nonumber \\
 & +\left(i\Delta_{k}/2-\Gamma_{-,k}/2\right)\left\langle aA_{\bar{r}g}^{k}\right\rangle -i\sum_{k'\neq k}\sqrt{2}g_{k'}\left\langle A_{gB}^{k'}A_{rg}^{k}\right\rangle \nonumber \\
 & +i\sqrt{2}g_{k}\left[\delta_{r,B}\left\langle a^{+}a\right\rangle \left\langle A_{gg}^{k}\right\rangle +\left(\left\langle a^{+}a\right\rangle +1\right)\left\langle A_{rB}^{k}\right\rangle \right].
\end{align}
The equation for the atom-atom correlation becomes
\begin{align}
 & \frac{\partial}{\partial t}\left\langle A_{gr}^{k}A_{r'g}^{k'}\right\rangle =\Bigl[i\left(\omega_{a}^{k'}-\omega_{a}^{k}\right)-\Lambda_{+,k}\nonumber \\
 & -\Gamma_{+,k}/2-\Lambda_{+,k'}-\Lambda_{+,k'}/2\Bigr]\left\langle A_{gr}^{k}A_{r'g}^{k'}\right\rangle \nonumber \\
 & -\left(i\Delta_{k}/2+\Gamma_{-,k}/2\right)\left\langle A_{g\bar{r}}^{k}A_{r'g}^{k'}\right\rangle \nonumber \\
 & +\left(i\Delta_{k'}/2-\Gamma_{-,k'}/2\right)\left\langle A_{gr}^{k}A_{\bar{r'}g}^{k'}\right\rangle \nonumber \\
 & -i\sqrt{2}g_{k}\Bigl(\delta_{r,B}\left\langle A_{gg}^{k}\right\rangle -\left\langle A_{Br}^{k}\right\rangle \Bigr)\left\langle aA_{r'g}^{k'}\right\rangle \nonumber \\
 & -i\sqrt{2}g_{k'}\left\langle aA_{rg}^{k}\right\rangle ^{*}\Bigl(\left\langle A_{r'B}^{k'}\right\rangle -\delta_{r',B}\left\langle A_{gg}^{k'}\right\rangle \Bigr).
\end{align}
The equation for the population and polarization of atoms become
\begin{align}
 & \frac{\partial}{\partial t}\left\langle A_{rr'}^{k}\right\rangle =-\left(i\Delta_{k}/2+\Gamma_{-,k}/2\right)\left\langle A_{r\bar{r'}}^{k}\right\rangle \nonumber \\
 & +\left(i\Delta_{k}/2-\Gamma_{-,k}/2\right)\left\langle A_{\bar{r}r'}^{k}\right\rangle -\Gamma_{+,k}\left\langle A_{rr'}^{k}\right\rangle \nonumber \\
 & -i\sqrt{2}g_{k}\Bigl(\delta_{r',B}\left\langle aA_{rg}^{k}\right\rangle -\delta_{r,B}\left\langle aA_{r'g}^{k}\right\rangle ^{*}\Bigr)\nonumber \\
 & +\Lambda_{+,k}\delta_{r',r}\left\langle A_{gg}^{k}\right\rangle +\Lambda_{-,k}\delta_{r',\bar{r}}\left\langle A_{gg}^{k}\right\rangle,
\end{align}
\begin{align}
 & \frac{\partial}{\partial t}\left\langle A_{gg}^{k}\right\rangle =-2\sqrt{2}g_{k}\mathrm{Im}\left\langle aA_{Bg}^{k}\right\rangle +2\Gamma_{-,k}\mathrm{Re}\left\langle A_{BD}^{k}\right\rangle \nonumber \\
 & -2\Lambda_{+,k}\left\langle A_{gg}^{k}\right\rangle +\Gamma_{+,k}\left(\left\langle A_{BB}^{k}\right\rangle +\left\langle A_{DD}^{k}\right\rangle \right).
\end{align}

To calculate the spectrum, we first solve the above
equations in the steady-state and then utilize the results as the
input parameters for the following simplified equations for the filter cavity-related
quantities:
\begin{align}
 &\frac{\partial}{\partial t}\left\langle b^{+}b\right\rangle =-\chi\left\langle b^{+}b\right\rangle +\beta2\mathrm{Im}\left\langle b^{+}a\right\rangle , \label{eq:bpb-DB}
 \end{align}
\begin{align}
 & \frac{\partial}{\partial t}\left\langle b^{+}a\right\rangle =\left[i\left(\omega_{f}-\omega_{c}\right)-\left(\chi+\kappa\right)/2\right]\left\langle b^{+}a\right\rangle \nonumber \\
 & -i\beta\left(\left\langle b^{+}b\right\rangle -\left\langle a^{+}a\right\rangle \right)-i\sum_{k}\sqrt{2}g_{k}\left\langle b^{+}A_{gB}^{k}\right\rangle,  \label{eq:bpa-DB}
\end{align}
\begin{align}
 & \frac{\partial}{\partial t}\left\langle b^{+}A_{gr}^{k}\right\rangle = -i\sqrt{2}g_{k}  \left\langle b^{+}a \right\rangle \left( \delta_{r,B} \left\langle A_{gg}^{k}\right\rangle -  \left\langle A_{Br}^{k}\right\rangle \right) \nonumber \\
 & + \left[i\left(\omega_{f}-\omega_{a}^{k}\right)-\chi/2-\Lambda_{+,k}-\Gamma_{+,k}/2\right]\left\langle b^{+}A_{gr}^{k}\right\rangle \nonumber \\
 & +i\beta\left\langle a^{+}A_{gr}^{k}\right\rangle - \left(i\Delta_{k}/2+\Gamma_{-,k}/2\right) \left\langle b^{+}A_{g\bar{r}}^{k}\right\rangle.   \label{eq:bpAk-DB}
\end{align}
In the last equation, we have utilized $\left\langle b^{+}aA^k_{BB}\right\rangle =\left\langle b^{+}a\right\rangle \left\langle A^k_{BB}\right\rangle $
and $\left\langle b^{+}aA^k_{gg}\right\rangle =\left\langle b^{+}a\right\rangle \left\langle A^k_{gg}\right\rangle $.

For $N$ identical atoms, we expect that  $\left\langle A^k_{gg}\right\rangle$,  $\left\langle aA^k_{rg}\right\rangle$, $\left\langle b^{+}A^k_{gr}\right\rangle $, $\left\langle A^k_{rr'}\right\rangle$  are identical for different atom $k$  and $\left\langle A^{k'}_{gr'}A^k_{rg}\right\rangle$ are identical for different atom pair $k',k$. As a result,  the number of independent elements is just $1+2+2+4+4+3 = 16$.

\section{Semi-analytical Expression of Spectrum Linewidth \label{sec:analytical-linewidth} }
In this section, we derive a semi-analytical expression for the steady-state spectrum linewidth.
To proceed, we consider the steady-state version of Eq. (\ref{eq:bpb-DB}) and (\ref{eq:bpAk-DB}) with $g=g_k$, $\omega_a=\omega^k_a$, $\Delta=\Delta_k$, $\Lambda_{\pm} = \Lambda_{\pm,k}$, $\Gamma_{\pm} = \Gamma_{\pm,k}$  for all $k$:

\begin{align}
& \left\langle b^{+}b\right\rangle =-i\left(\beta/\chi\right)\left(\left\langle b^{+}a\right\rangle -\left\langle ba^{+}\right\rangle \right), \label{eq:bpb-sol} \\
& \xi\left\langle b^{+}A^k_{gB}\right\rangle -i\frac{1}{2}\left(i\Delta+\Gamma_{-}\right)\left\langle b^{+}A^k_{gD}\right\rangle  \nonumber \\
&=\beta\left\langle a^{+}A^k_{gB}\right\rangle +\sqrt{2}g\left(\left\langle A^k_{BB}\right\rangle -\left\langle A^k_{gg}\right\rangle \right)\left\langle b^{+}a\right\rangle , \label{eq:bpAB} \\
& \xi\left\langle b^{+}A^k_{gD}\right\rangle -i\frac{1}{2}\left(i\Delta+\Gamma_{-}\right)\left\langle b^{+}A_{gB}\right\rangle  \nonumber \\
& =\beta\left\langle a^{+}A^k_{gD}\right\rangle + \sqrt{2} g \left\langle A^k_{BD}\right\rangle \left\langle b^{+}a\right\rangle . \label{eq:bpAD}
\end{align}
Here, we have introduced $ \xi=\omega_{a}-\omega_{f}-i\left(\chi/2+\Lambda_{+}+\Gamma_{+}/2\right)$. We solve Eqs. (\ref{eq:bpAB}) and (\ref{eq:bpAD}) and obtain
\begin{equation}
\left\langle b^{+}A^k_{gB}\right\rangle =\beta\varepsilon_{1}+\varepsilon_{2}\left\langle b^{+}a\right\rangle  \label{eq: bpAB-sol}
\end{equation}
with the abbreviations $ \varepsilon_{1}=\varepsilon[\xi\left\langle a^{+}A^k_{gB}\right\rangle +\frac{1}{2}i\left(i\Delta+\Gamma_{-}\right)\left\langle a^{+}A^k_{gD}\right\rangle ]$, $\varepsilon_{2}=\varepsilon \sqrt{2} g [\xi (\left\langle A^k_{BB}\right\rangle -\left\langle A^k_{gg}\right\rangle ) +\frac{1}{2}i\left(i\Delta+\Gamma_{-}\right)  \left\langle  A^k_{BD}\right\rangle]$  and $ \varepsilon^{-1}= \xi^{2}+\frac{1}{4}\left(i\Delta+\Gamma_{-}\right)^{2}$.  Then,  we consider the steady-state version of Eq. (\ref{eq:bpa-DB}) and its complex conjugate:
\begin{align}
	&0=\left[i\left(\omega_{f}-\omega_{c}\right)-\left(\chi+\kappa\right)/2\right]\left\langle b^{+}a\right\rangle \nonumber \\
	&-i\beta\left(\left\langle b^{+}b\right\rangle -\left\langle a^{+}a\right\rangle \right)-iN\sqrt{2}g\left\langle b^{+}A^k_{gB}\right\rangle, \\
	&0=-\left[i\left(\omega_{f}-\omega_{c}\right)+\left(\chi+\kappa\right)/2\right]\left\langle ba^{+}\right\rangle \nonumber \\
	&+i\beta\left(\left\langle b^{+}b\right\rangle -\left\langle a^{+}a\right\rangle \right)+iN\sqrt{2}g\left\langle b^{+}A^k_{gB}\right\rangle^{*}.
\end{align}
Inserting Eqs. (\ref{eq:bpb-sol}) and (\ref{eq: bpAB-sol}) into the above equations, we obtain the coupled equations
\begin{equation}
\left[\begin{array}{cc}
\tau^{*} & -\beta^{2}/\chi\\
-\beta^{2}/\chi & \tau
\end{array}\right]\left[\begin{array}{c}
\left\langle b^{+}a\right\rangle \\
\left\langle ba^{+}\right\rangle
\end{array}\right]=i\beta\left[\begin{array}{c}
\nu^{*}\\
-\nu
\end{array}\right]
\end{equation}
with the abbreviations $ \tau=i(\omega_{f}-\omega_{c}-N\sqrt{2}g\varepsilon_{2}^{*})+\left(\chi+\kappa\right)/2+\beta^{2}/\chi$ and $\nu=\left\langle a^{+}a\right\rangle -N\sqrt{2}g\varepsilon_{1}^{*} $. The solution is
\begin{align}
\left\langle b^{+}a\right\rangle &=\frac{i\beta\left(\nu^{*}\tau-\nu\beta^{2}/\chi\right)}{\tau\tau^{*}-\beta^{4}/\chi^{2}}, \\\left\langle ba^{+}\right\rangle &=\frac{i\beta\left(\nu^{*}\beta^{2}/\chi-\tau^{*}\nu\right)}{\tau\tau^{*}-\beta^{4}/\chi^{2}},
\end{align}
and  Eq. (\ref{eq:bpb-sol}) yields
\begin{equation}
\left\langle b^{+}b\right\rangle =\frac{\beta^{2}}{\chi}\frac{\nu^{*}\tau+\tau^{*}\nu-\left(\nu+\nu^{*}\right)\beta^{2}/\chi}{\tau\tau^{*}-\beta^{4}/\chi^{2}}. \label{eq:bpb-ana}
\end{equation}

To compute the spectrum, we require that $\beta$ is small and thus the above expression can be simplified as
\begin{equation}
\left\langle b^{+}b\right\rangle \approx\frac{\beta^{2}}{\chi}2\mathrm{Re}\frac{\nu}{\tau}.
\end{equation}
This expression suggests that the linewidth of the spectrum is mainly determined by $\tau$.  Ignoring the terms proportional to $\beta$ again, we can approximate this term as
\begin{equation}
\tau\approx i\left(\omega_{f}-\omega_{c}-Z\right)+\kappa/2 \label{eq:tau-app}
\end{equation}
 with the abbreviation
\begin{align}
& Z=\frac{N2g^{2}}{\left(\xi^{*}\right)^{2}+\left(i\Delta-\Gamma_{-}\right)^{2}/4}
 [  \xi^{*}\left( \left\langle A^k_{BB}\right\rangle -\left\langle A^k_{gg}\right\rangle \right) \nonumber \\
& -i\frac{1}{2}(-i\Delta + \Lambda_{-}) \left\langle A^k_{DB}\right\rangle] , \label{eq:Zfun}
\end{align}
and the parameter $\xi\approx\omega_{a}-\omega_{f}-i\left(\Lambda_{+}+\Gamma_{+}/2\right)$.

To proceed further, we consider the resonant condition $\omega_f \approx \omega_a^k+i \Gamma/2 \approx \omega_c+i \Gamma/2$ (with the linewidth $\Gamma$ to be determined) and
$\Lambda_{-}=0$ (i.e. the balanced pumping $\eta_{+,k}=\eta_{-,k}$), which allows us to simplify Eq. (\ref{eq:Zfun}) as
\begin{align}
& Z  =- \frac{i2Ng^{2}}{\left(\Gamma/2 + \Lambda_{+}+\Gamma_{+}/2\right)^{2}+\left(\Delta\right)^{2}/4} [ i(\Delta/2) \left\langle A^k_{DB}\right\rangle  \nonumber \\
& + \left(\Gamma/2 + \Lambda_{+}+\Gamma_{+}/2\right) \left(\left\langle A^k_{BB}\right\rangle -\left\langle A^k_{gg}\right\rangle \right)]. \label{eq:Gamma}
\end{align}
Inserting the above expression to Eq. (\ref{eq:tau-app}), we establish the relation $ {\rm Re}\tau \approx - \Gamma/2 + {\rm Im}Z +\kappa/2$.  Equating this expression to $\Gamma/2$, we obtain an equation for $\Gamma$, which can be solved numerically. If the spectrum linewidth is small, i.e. $\Gamma <\Lambda_{+},\Delta$, we can simplify Eq. (\ref{eq:Gamma}) by ignoring the dependence of the denominator on  $\Gamma$. In this case, we can achieve the following semi-analytical expression for the linewidth
\begin{align}
& \Gamma  = \left[1+\theta (\left\langle A^k_{BB}\right\rangle-\left\langle A^k_{gg}\right\rangle)/2\right]^{-1} \{
\kappa/2  -\theta[ (\Lambda_{+} \nonumber \\
&+\Gamma_{+}/2 ) \left(\left\langle A^k_{BB}\right\rangle -\left\langle A^k_{gg}\right\rangle \right)  - (\Delta/2){\rm Im}\left\langle A^k_{DB}\right\rangle ]\} \label{eq:linewidth}
\end{align}
with the abbreviation $\theta = 2Ng^2/[\left( \Lambda_{+}+\Gamma_{+}/2\right)^2 + \Delta^2/4] $, which can be rewritten as Eq. (5) in the main text. The expression (\ref{eq:linewidth}) indicates that the linewidth is determined by not only the  population inversion of the bright state,  i.e. $\left\langle A^k_{BB}\right\rangle - \left\langle A^k_{gg}\right\rangle$,  but also the off-diagonal element $\left\langle A^k_{DB}\right\rangle$.  It is through these elements the dark atomic excited states contribute to the lasing.

 \begin{figure}
\begin{centering}
\includegraphics[scale=0.30]{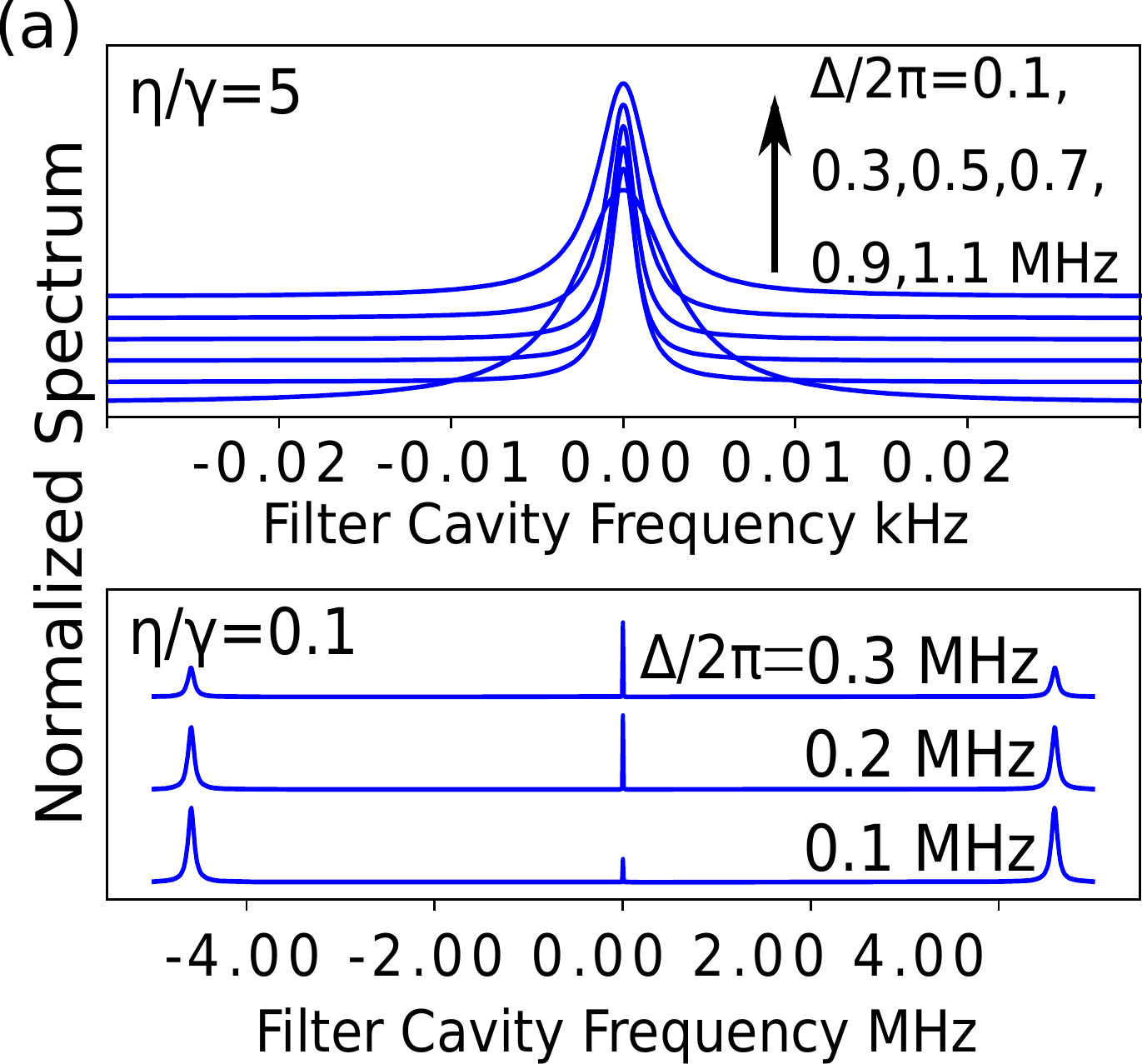}
\includegraphics[scale=0.32]{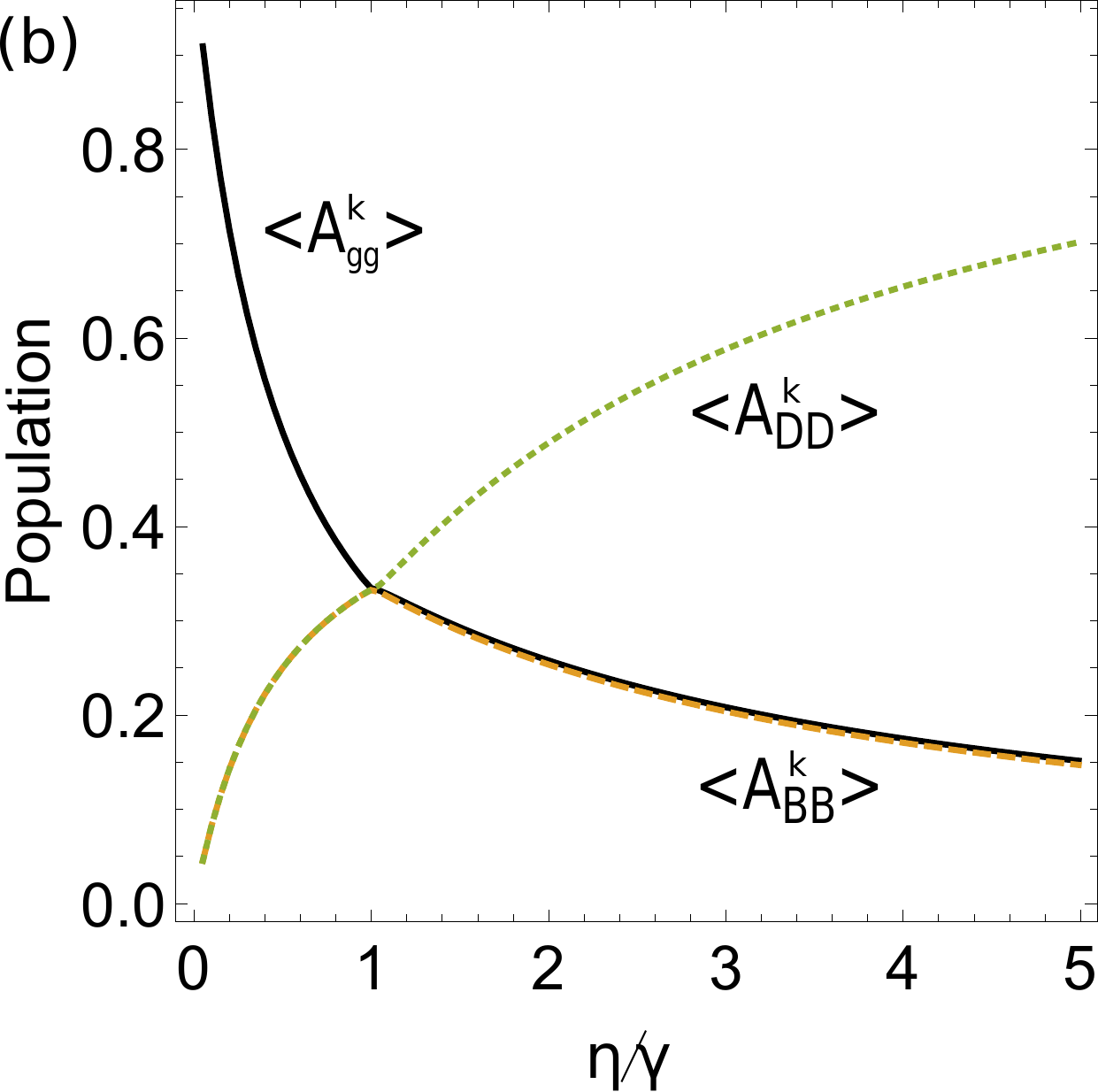}
\par\end{centering}
\begin{centering}
\includegraphics[scale=0.32]{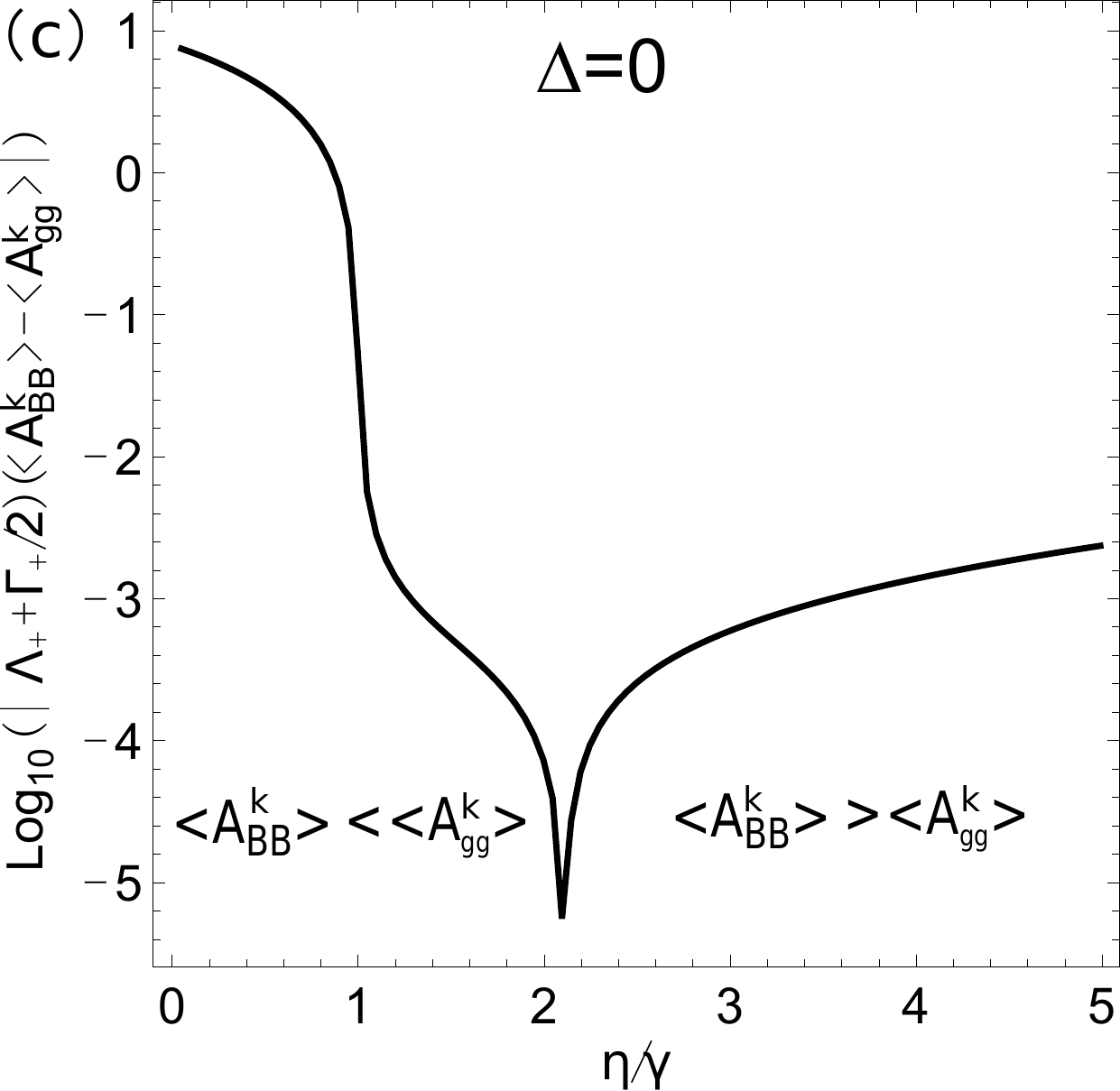}
\includegraphics[scale=0.34]{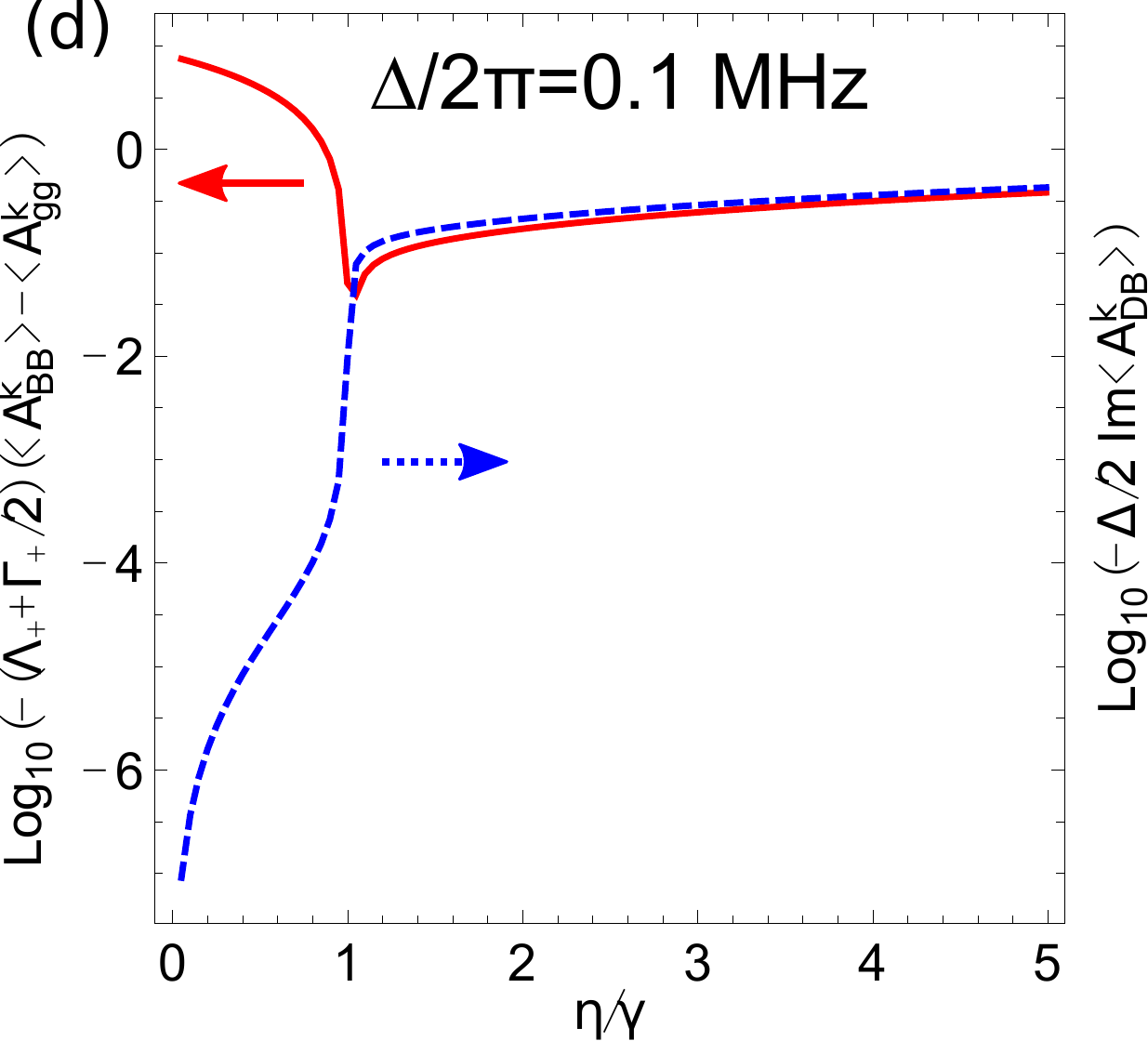}
\par\end{centering}
\centering{}\caption{\label{fig:linewidth-DB} Supplemental results  for systems with $2.5\times10^5$ atoms under incoherent pumping. Panel (a) shows the spectra with increasing Zeeman splitting $\Delta$ for weak pumping $\eta=0.1\gamma$ (lower part) and strong pumping  $\eta=5\gamma$ (upper part). Panel (b) shows the populations $\left\langle  A^k_{gg} \right\rangle $ (black solid curve),  $\left\langle A^k_{BB} \right\rangle $(red dashed curve) and  $\left\langle  A^k_{DD} \right\rangle $ (green dotted curve), which are similar for the systems with and without Zeeman splitting $\Delta$.   Panel (c) shows the population difference between the bright atomic excited state $\left\langle A^k_{BB} \right\rangle $ and the ground state  $\left\langle A^k_{gg} \right\rangle $ (multiplied with a factor), which is negative for $\eta/\gamma<2.2$ but becomes positive for $\eta/\gamma>2.2$, for the system without Zeeman splitting.  Panel (d) shows the population difference as in (c) and the imaginary part of the coherence between the bright and dark atomic excited state $\left\langle A^k_{DB} \right\rangle$ (multiplied with a factor) for the system with Zeeman splitting.  Other parameters are the same as used in the main text. }
\end{figure}

Figure \ref{fig:linewidth-DB} supplements the results shown in  Fig. 3 in the main text for $2.5\times 10^5$ atoms in the absence and presence of the magnetic field-induced Zeeman splitting $\Delta$. Fig. \ref{fig:linewidth-DB}(a) shows  the influence of the Zeeman-splitting on the steady-state spectrum. For weak pumping $\eta=0.1\gamma$ the two side peaks become weaker while the center peak gets stronger with increasing $\Delta$ due to the increased influence of the dark atomic excited state. For strong pumping $\eta=5\gamma$, the single peak becomes even narrower when $\Delta$ increases from $0.1$ MHz to $0.3$ MHz, but it widens again when $\Delta$ increases further. Fig. \ref{fig:linewidth-DB}(b) shows that the population of the ground state $\left\langle  A^k_{gg} \right\rangle $ and dark atomic excited state $\left\langle  A^k_{DD} \right\rangle $ decrease and increase monotonously with increased pumping $\eta$, while the population of the bright atomic excited state $\left\langle  A^k_{BB} \right\rangle $ first increases for $\eta$ smaller than the spontaneous emission rate $\gamma$ and then decreases, because the stimulated absorption balances the emission. The change of population is similar for the systems with $\Delta=0$ and $\Delta \neq 0$, and the linewidth reduction cannot be simply explained by the population dynamics.

To proceed, we compare the population difference $\left\langle  A^k_{BB} \right\rangle - \left\langle  A^k_{gg} \right\rangle$ between the bright atomic excited state and the ground state with the coherence $\left\langle  A^k_{DB} \right\rangle$ between the bright and dark atomic excited states, see Fig. \ref{fig:linewidth-DB}(c) and (d). Fig. \ref{fig:linewidth-DB}(c) shows that, in the absence of  Zeeman splitting $\Delta =0$, the coherence is always zero and the population difference is negative for the pumping $\eta$ smaller than $2.2\gamma$ but becomes positive otherwise. Thus, in this case, the linewidth narrowing can be attributed to the population inversion as in normal lasing.  However, Fig. \ref{fig:linewidth-DB}(d) shows that, in the presence of the Zeeman-splitting $\Delta/2\pi =0.1$ MHz, the population difference is always negative, while the imaginary part of the coherence becomes much larger than the population difference for large pumping $\eta$. Associated with the expression (\ref{eq:linewidth}), these results indicate that in the case with  $\Delta \neq 0$ the negative population difference causes light absorption and spectral broadening while the coherence is responsible for the gain and the spectrum narrowing.

\section{Bright and Dark Pseudo-Dicke States \label{sec:Dicke}}
We have further examined how the higher excited states of the atomic ensemble are involved in the spectral features. For two-level atoms, these states can be well characterized by the Dicke states \citep{RHDick} $\left| J,M \right\rangle$, where the collective spin quantum number $J\leq N/2$ characterizes the symmetry of Dicke states and the collective coupling of these states to the quantized field,  and $-J \leq M \leq J$ describes the excitation of the atoms. According to our previous study \citep{KDebnath}, the mean of these numbers can be calculated with the quantities in the second-order mean field theory:  $M=\frac{1}{2}  N ( \left\langle A_{ee}^k \right\rangle - \left\langle A_{gg}^k \right\rangle)$ and $J^2=\frac{3}{4} N +  N(N-1) \left[ \left\langle A_{ge}^k A_{eg}^{k'} \right\rangle + \frac{1}{4} \left\langle (A_{ee}^k - A_{gg}^k) (A_{ee}^{k'} - A_{gg}^{k'})  \right\rangle\right]$  ($k\neq k'$)  for identical atoms. Here, $e,g$ indicates the upper and lower level of the two-level atoms.

To apply the Dicke states to our three-level atoms, we consider the two transitions $D_k \to g_k$, $B_k \to g_k$  as shown in Fig. 1 (c) in the main text as two sub-two-level systems and define the pseudo Dicke states $\left| J_s,M_s \right\rangle$ with these transitions. Then,  the quantum numbers $J_s,M_s$ can be computed by simply replacing $e$ with $D$ or $B$ in the above formulas. We refer to these states as pseudo Dicke states since the two transitions are not independent but connected through conservation of population, i.e. $\left\langle A_{DD}^k \right\rangle + \left\langle A_{BB}^k \right\rangle + \left\langle A_{gg}^k \right\rangle =1$.

Fig. 4(a) in the main text shows the evolution of the Dicke quantum numbers in the Dicke ladders as  the pumping increases.  The evolution starts from the bottom-right corner along the bottom boundary and explores the sub-radiative states at the left corner for the bright transition while it continues further along the upper boundary and explores the  symmetric Dicke states close to the upper-right corner for the dark transition.  The evolution is similar for the systems  without and with Zeeman splitting $\Delta$. However, the careful examination shows $M_B>0$ ( $M_B<0$) for  $\Delta=0$ ( $\Delta\neq 0$).

By assembling the information revealed by Fig. \ref{fig:linewidth-DB} and Fig. 4(a) in the main text , we conclude that for weak pumping the spectrum can be associated with the dressed atom-light states formed by weakly excited states of the atomic ensemble.  For stronger pumping, we expect that the super-narrow spectrum is associated with dressed atom-light states formed by highly excited states of the atomic ensemble. It is difficult to give explicit expressions for such dressed states but we expect that similar energy diagrams as shown in Fig. \ref{fig:dressed-states} can be formed by replacing the ground and first excited state of the atomic ensemble with the adjacent highly excited states of the atomic ensemble and thus expect still three peaks in the spectrum for strong pumping. By reexamining the spectra on the logarithmic scale, see Fig. 4(b) in the main text , we do see two other side peaks beside the super-narrow center peak. Since these peaks are orders of magnitude smaller than the center peak, we can not see them in normal scale.

 \begin{figure}[t!]
\begin{centering}
\includegraphics[scale=0.35]{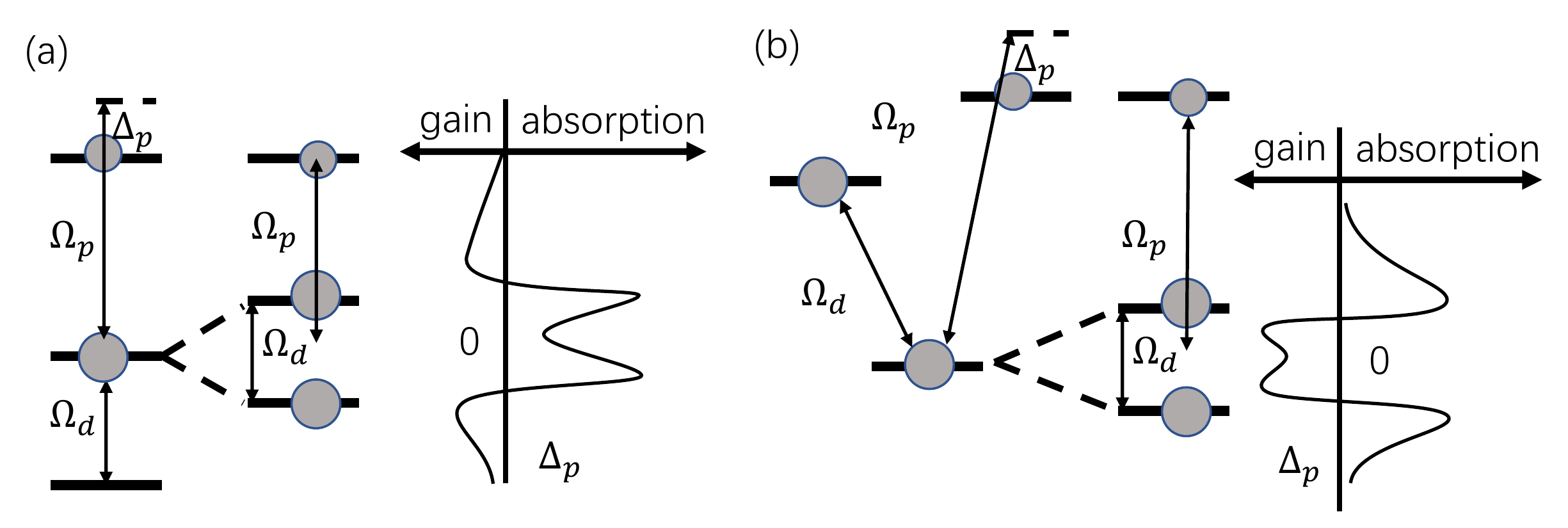}
\par\end{centering}
\centering{}\caption{\label{fig:LWI} Lasing without inversion from systems with cascade (a) and  V-type (b) energy scheme (left), where $\Omega_d$,$\Omega_p$ denote the coupling with the driving and probe classical field, respectively, $\Delta_p$ is the frequency detuning of the probe field. The middle part of both panels shows the formation of the  dressed states due to the coupling with the driving field. The right part shows the gain and absorption for different detuning $\Delta_p$. The spheres indicate the relative population of the levels. Figure redrawn from \citep{JMompart} with the permission of J. Mompart. }
\end{figure}

\section{Comparison with Lasing without Inversion \label{sec:lwi}}
The lasing mechanism revealed by our analysis  bears much resemblance with lasing without inversion (LWI) \citep{JMompart}, which also relies on atomic coherence in the absence of population inversion.
Some LWI schemes rely on atoms interacting strongly with a driving field and weakly with a probe field and do not necessarily require incoherent pumping.  Fig. \ref{fig:LWI} shows cascade (a) and V-type (b) three-level systems for LWI \citep{JMompart}, which are similar to our systems, both in the bare state and the dressed eigenstate basis due to the coupling with the strong driving field. The rightmost panels show the amplification and absorption of the probe field as function of the detuning $\Delta_p$ (right). 

Figure 1(c) in our main text is similar to Fig.  \ref{fig:LWI} (a) except that the coupling with the driving field is replaced by the collective atom-cavity mode coupling, the coupling with the probe field is replaced by the Zeeman splitting, and the upper level is shifted to coincide with the middle level. As shown in Fig. \ref{fig:dressed-states}, we can introduce the dressed states due to the atom-cavity coupling, and understand the influence of the Zeeman splitting as a perturbation, in the same way as the middle part of Fig.  \ref{fig:LWI} (a).  However, in our system, the perturbation introduces another kind of dressed states  which are occupied due to the incoherent pumping and are responsible for the super-narrow lasing. In addition, Fig. 1 (b) in our main text (ignoring the center excited level) is similar to Fig. \ref{fig:LWI} (b)  except that the driving and probe field are replaced by the same cavity field with identical coupling strength. 

In some studies on LWI incoherent pumping process  \citep{YZhu} and some effects of collective atom-light interactions \citep{GYang} are included, and we believe that unification of those ideas with the ones of the present work may lead to further understanding and proposals for new and even better schemes.


\begin{thebibliography}{1}
\bibitem{PDick} The collective dynamics of an ensemble of $N$ identical two-level atoms can be described by Dicke states $\left|J,M\right \rangle$, where the collective spin quantum number $J \leq N/2$ describes the symmetry and collective coupling of the states to the quantized field, and $-J \leq M \leq J$ describes the excitation of the system. As shown in \citep{YZhang1,NShammah} and in Fig. \ref{fig:DickePicture}(a), for $\eta<\gamma$ the incoherent pumping mainly causes reduction of the Dicke state quantum number $J$ while for $\eta>\gamma$, it mainly increases $J$ and hence the effective coupling strength.
\end{thebibliography}
\end{document}